\RequirePackage{fix-cm}
\documentclass[smallextended]{svjour3}       
\smartqed  
\usepackage[british]{babel}
\usepackage{graphicx}
\usepackage{amsmath}
\usepackage{amssymb}

\usepackage{mathptmx}      
\journalname{Journal of Statistical Physics}
\begin{document}

\title{Is the Boltzmann equation reversible? A large deviation perspective
on the irreversibility paradox.
}

\titlerunning{A large deviation perspective
on the irreversibility paradox and the Boltzmann equation}        

\author{Freddy Bouchet 
}


\institute{F. Bouchet \at
              Univ Lyon, Ens de Lyon, Univ Claude Bernard, CNRS, Laboratoire de Physique, Lyon, France \\
              \email{freddy.bouchet@ens-lyon.fr}           
}

\date{Received: date / Accepted: date}

\maketitle

\begin{abstract}
We consider the kinetic theory of dilute gases in the Boltzmann\textendash Grad
limit. We propose a new perspective based on a large deviation estimate
for the probability of the empirical distribution dynamics. Assuming
Boltzmann molecular chaos hypothesis (Stosszahlansatz), we derive
a large deviation rate function, or action, that describes the stochastic process for the empirical
distribution. The quasipotential for this action is the negative of the
entropy, as should be expected. While the Boltzmann equation appears
as the most probable evolution, corresponding to a law of large numbers,
the action describes a genuine reversible stochastic process for the
empirical distribution, in agreement with the microscopic reversibility.
As a consequence, this large deviation perspective gives the expected
meaning to the Boltzmann equation and explains its irreversibility
as the natural consequence of limiting the physical description to
the most probable evolution. More interestingly, it also quantifies
the probability of any dynamical evolution departing from solutions
of the Boltzmann equation. This picture is fully compatible with the
heuristic classical view of irreversibility, but makes it much more precise in various
ways. We also explain that
this large deviation action provides a natural gradient structure
for the Boltzmann equation. 
\end{abstract}

\section{Introduction\label{sec:Introduction-1}}

The Boltzmann equation \cite{boltzmann1896vorlesungen} (see \cite{brush2013kinetic,brush2016kinetic}
for an english translation) is a cornerstone of statistical physics.
It describes dilute gas dynamics at a macroscopic level, and has been
the starting point for the kinetic theory of many other physical phenomena:
the derivation of hydrodynamic equations \cite{chapman1990mathematical,saint2009hydrodynamic},
the kinetic theory of self gravitating systems \cite{Binney_Tremaine_1987_Galactic_Dynamics},
the relativistic Boltzmann equation, lattice Boltzmann algorithms
for fluid mechanics \cite{frisch1986lattice}, nuclear physics, and
so on. While the underlying microscopic Hamiltonian dynamics is time
reversible, Boltzmann\textquoteright s equation increases the entropy,
as proven by Boltzmann\textquoteright s H-theorem. This irreversibility
paradox has played a crucial role in the early development of statistical
physics and led to long controversies, for instance between Boltzmann
and Zermelo, that involved many of the leading physicists and mathematicians
of the late nineteen century (see \cite{brush2013kinetic,brush2016kinetic}
for a collection of basic papers from the second half of the nineteenth
century on the subject). This apparent paradox and the fact that
the irreversible evolution of macroscopic laws is a natural consequence
of the reversible microscopic dynamics was already well understood
by Thomson \cite{thomson18759}, Maxwell and Boltzmann \cite{boltzmann1896entgegnung}.
``Boltzmann's thoughts on this question have withstood the test of
time'', as stated by Lebowitz in a very nice discussion \cite{lebowitz1993boltzmann}
(see also \cite{lebowitz2008time,Goldstein_Lebowitz_2004PhyD..193...53G}
for a pedagogical discussion of the irreversibility paradox and the
meaning of entropy in relation with the Boltzmann equation). Boltzmann's
explanation of the irreversibility paradox can be read in classical
physics books \cite{feynman1967character} or more specialized mathematical
physics books \cite{cercignani1988boltzmann,Spohn_1991}.

In a nutshell, Boltzmann\textquoteright s explanation is that the
Boltzmann equation describes at a macroscopic level not all but most
of the microscopic evolutions. If we consider a set of microscopic
initial conditions compatible with a macroscopic distribution $f_{0}(\mathbf{r},\mathbf{v})$,
then most of these initial conditions will have a dynamics compatible
with the solution of the Boltzmann equation $f(\mathbf{r},\mathbf{v},t)$
with initial conditions $f(\mathbf{r},\mathbf{v},t)=f_{0}(\mathbf{r},\mathbf{v})$.
More precisely, if we denote $\epsilon$ the inverse of the number
of particles which are contained in a volume equal to the cube of
the mean free path $l$ ($\epsilon=1/\rho l^{3}$where $\rho$ is
the gas density), then the number of microscopic initial conditions
which are compatible with $f_{0}(\mathbf{r},\mathbf{v})$ and that
will actually follow the solution of the Boltzmann equation increases
exponentially with $1/\epsilon$. The evolution of the actual microscopic
dilute gas should thus be observed with a dynamics that follows the
solution of the Boltzmann equation with an overwhelming probability.
Three ingredients are key in this explanation \cite{lebowitz1993boltzmann}:
a) the great disparity between microscopic and macroscopic scales,
measured by the smallness of the parameter $\epsilon$, b) the fact
that events are, as put by Boltzmann, determined not only by differential
equations but also by initial conditions, and c) the use of probabilistic
reasoning: it is not every microscopic state of a macroscopic system
that will evolve in accordance with the second law, but only the `majority''
of states a majority which however becomes so overwhelming when the
number of atoms in the system becomes very large that irreversible
behavior becomes a near certainty. 

While this heuristic understanding of the irreversibility paradox
makes consensus, the actual proof of a theorem or even a precise theoretical
characterization and quantification of this statement, are still lacking.
This issue led to a very difficult mathematical challenge: proving
the validity of the Boltzmann equation from the microscopic Hamiltonian
dynamics. A first step has been achieved by Lanford\textquoteright s
proof in the 70' \cite{lanford1975time}, valid for times shorter
than the mean collision time. While several impressive improvements
have been achieved recently, for instance proofs for other potentials
than hard sphere interactions \cite{gallagher2014newton,pulvirenti2014validity},
the time interval for which those theorem have been proven remains
quite small. \\

In this article, we will convey the idea that, although fascinating
and very interesting, Lanford\textquoteright s proof and successive
mathematical developments should be complemented by a conceptually
different approach. We propose a perspective based on a large deviation
theory for the estimate for the probability of the empirical distribution
dynamics. Making assumptions akin to Boltzmann ones (mainly Boltzmann
molecular chaos hypothesis), we derive a large deviation rate function,
or action, that describes asymptotically for small $\epsilon$ (or
equivalently the Boltzmann\textendash Grad limit), the stochastic
process for the empirical distribution. We consider the dynamics of
the empirical distribution $f_{\epsilon}(\mathbf{r},\mathbf{v},t)\equiv\epsilon\sum_{n=1}^{N}\delta\left(\mathbf{v}-\mathbf{v}_{n}(t)\right)\delta\left(\mathbf{r}-\mathbf{r}_{n}(t)\right)$.
Our main result is that the probability that $\left\{ f_{\epsilon}(t)\right\} _{0\leq t\leq T}$
remains in a neighborhood of a prescribed path $\left\{ f(t)\right\} _{0\leq t\leq T}$
verifies 
\begin{equation}
\mathbb{P}\left[\left\{ f_{\epsilon}(t)\right\} _{0\leq t<T}=\left\{ f(t)\right\} _{0\leq t<T}\right]\underset{\epsilon\downarrow0}{\asymp}\exp\left(-\frac{\int_{0}^{T}\mbox{d}t\,L\left[f,\dot{f}\right]}{\epsilon}\right),\label{eq:Large_Deviation_Action}
\end{equation}
where $\underset{\epsilon\downarrow0}{\asymp}$ mean a log-equivalence
when $\epsilon$ goes to zero, and where the large deviation Lagrangian
$L$ can be computed through a Legendre\textendash Fenchel transform
from the Hamiltonian $H$ given by $H\left[f,p\right]=H_{R}\left[f,p\right]+H_{C}\left[f,p\right]$
with 
\begin{multline}
H_{R}\left[f,p\right]=\frac{1}{2}\int\mbox{d}\mathbf{v}{}_{1}\mbox{d}\mathbf{v}{}_{2}\mbox{d}\mathbf{v}'_{1}\mbox{d}\mathbf{v}'_{2}\mbox{d}\mathbf{r}\,w(\mathbf{v}'_{1},\mathbf{v}'_{2};\mathbf{v}_{1},\mathbf{v}_{2})\\
\times f(\mathbf{r},\mathbf{\mathbf{v}_{1})}f\left(\mathbf{r},\mathbf{v}_{2}\right)\left\{ \mbox{e}^{\left[-p\left(\mathbf{r},\mathbf{v}_{1}\right)-p\left(\mathbf{r},\mathbf{v}_{2}\right)+p\left(\mathbf{r},\mathbf{v}'_{1}\right)+p\left(\mathbf{r},\mathbf{v}'_{2}\right)\right]}-1\right\} \label{eq:Hamiltonian-Collisions-1}
\end{multline}
and 
\begin{equation}
H_{C}\left[f,p\right]=-\int\mbox{d}\mathbf{r}\mbox{d}\mathbf{v}\,p(\mathbf{r},\mathbf{v})\mathbf{v}.\frac{\partial f}{\partial\mathbf{r}}(\mathbf{r},\mathbf{v}).\label{eq:Hamiltonian-Transport-1}
\end{equation}
Please note that the large deviation Hamiltonian $H$ is a probabilistic concept and is not the Hamiltonian of the microscopic dynamics, in the analytical mechanics theory. We note that the result (\ref{eq:Large_Deviation_Action}) is much in the spirit of large deviation results for the macroscopic description of paths for Markov processes \cite{Derrida_Lebowitz_1998_PhRvL} or of macroscopic fluctuation theory \cite{bertini2015macroscopic}, in other contexts.

We prove that this action $\int_{0}^{T}\mbox{d}t\,L\left[f,\dot{f}\right]$
is such that the Boltzmann equation appears as the most probable evolution.
As a consequence, this gives the expected meaning to the Boltzmann
equation as being the most probable evolution and a law of large numbers
in the limit $\epsilon\rightarrow0$. However, beyond this law of
large numbers, the probability of any macroscopic paths $\left\{ f(t)\right\} _{0\leq t\leq T}$
departing from the solutions of the Boltzmann equation is fully quantified
by (\ref{eq:Large_Deviation_Action}). The concentration of the path
measure is clearly quantified by $\epsilon$. We prove that the entropy
$S[f]=-\int\mbox{d}\mathbf{v}\mbox{d}\mathbf{r}\,f\log(f)$ is the
negative of the quasi-potential of the stochastic process for the
empirical density, for any $f$ with the proper mass, energy and momentum,
as should be expected. 

In order to discuss the irreversibility/reversibility paradox it is useful to use two distinct and related notions of time reversibility. Time reversibility for a dynamical system, applied to the mechanical system of particles, states that under change of the sign of time and velocity reversal, one obtains again a solution of the dynamical system equations. Time reversibility for a stochastic process, states that backward and forward histories of the stochastic process, also up to velocity reversal, have the same probabilities. The definition of the time reversibility of a stochastic process and its relation with detailed balanced are precisely discussed in section \ref{subsec:Quasipotential,-Hamilton-Jacobi} point 9 (without velocity inversion) and point 10 (with velocity inversion). The key point we want to stress is that the time reversibility of the microscopic dynamics and the consideration of the microcanonical measure at a fixed energy, necessarily imply the time reversibility of the stochastic process of the empirical density. As explained in section \ref{subsec:Quasipotential,-Hamilton-Jacobi} point 9 (without velocity inversion) and point 10, the time reversibility of the stochastic process for the empirical density has to translate into a time-reversal symmetry for the action. In section \ref{Time_reversal_symmetry_Boltzmann}, we check this property as a time-reversal symmetry property of both the Lagrangian $L$
and the Hamiltonian $H$.  The main conclusion is that the action $\int_{0}^{T}L\ \mbox{d}t$ is time reversible, and quantifies at a large deviation level the probabilities of the macroscopic evolution. Macroscopic dynamics is itself a time reversible stochastic process as one should expect. Moreover, this gives its full dynamical meaning to
the entropy, in relation with recent fluctuation theorems. Those results
are thus fully compatible with the classical picture of the irreversibility paradox but extend it
and clarify it in several ways.  Moreover from the
action (\ref{eq:Large_Deviation_Action}), the probability for the
evolution of any macroscopic variable can be computed. 

One clarification is worth stressing. For instance, we obtain the
large deviation result (\ref{eq:Large_Deviation_Action}) as a consequence
of the molecular chaos hypothesis (Stosszahlansatz), and the
action is actually time reversible. As a consequence we conclude that
the irreversibility is not a consequence of the molecular
chaos hypothesis (Stosszahlansatz); but it is rather a consequence
of describing only the most probable evolution of the empirical density,
or equivalently the evolution of the averaged evolution of the empirical
density (law of large numbers).\\

This paper is not a mathematical one. We will actually not be able
to derive the action that describes the large deviations of the empirical
distribution dynamics from the microscopic Hamiltonian dynamics. We
will achieve a much less ambitious goal. Starting from the classical
Boltzmann\textquoteright s molecular chaos hypothesis (Stosszahlansatz),
not approximating the evolution of the empirical density by the evolution
of its average (law of large numbers) but looking at the stochastic
process of all possible effects of molecule collisions, we will derive
the action that describes the large deviations of the empirical distribution
dynamics. With the same spirit and hypothesis as Boltzmann\textquoteright s
one, we will extend his results from the law of large number level
(the Boltzmann equation), to the large deviations level. Although
the writing of the paper is at the level of rigor found in any theoretical
physics textbook, the fact that the large deviation action is a consequence
of Boltzmann's molecular chaos hypothesis should be considered as
clear and rigorous. This argument is worth as much as Boltzmann's
molecular chaos hypothesis is believed to be natural. However it is
also clear that this is not a derivation from Newton's law, and such
a derivation would require to prove the validity of Boltzmann's molecular
chaos hypothesis, or an actual proof along other routes. 

Proving either the Boltzmann equation and/or the validity of this
natural large deviation action remains an open challenge for the future.
We note however that Rezakhanlou has proven \cite{rezakhanlou1998large}
a large deviation result for 1D stochastic dynamics mimicking the
hard sphere dynamics. Rezakhanlou action is actually the same as the
one we deduce from Boltzmann's molecular chaos hypothesis. This rigorous
result, and the fact that Boltzmann's molecular chaos is an extremely
natural hypothesis are clear hints that the formula (\ref{eq:Large_Deviation_Action}-\ref{eq:Hamiltonian-Transport-1})
actually describes the large deviations of any generic dilute gas
dynamics. Moreover, after the first writing of this article, for hard spheres and in the Boltzmann-Grad limit, Bodineau, Gallagher, Saint-Raymond and Simonella \cite{bodineau2020fluctuation} have derived large deviation asymptotics that give an information equivalent to the large deviation formula (\ref{eq:Large_Deviation_Action}-\ref{eq:Hamiltonian-Transport-1}), and valid for times of order of one collision time. Those impressive results extend for large deviations, Lanford's type theorems for the law of large number. 

We also prove in this paper that this large deviation action provides
a natural gradient structure for the Boltzmann equation. The Boltzmann
collision operator is the gradient of the entropy in a generalized
sense, where the measure of the distance to define the gradient is
related to the large deviation action. The transport term of the Boltzmann
equation is transverse to the gradient. This gradient structure has
a very natural physical interpretation and might have very interesting
mathematical consequences. \\

The original contributions of this work are: a) the derivation of the Boltzmann large deviation action from the molecular chaos hypothesis, giving an easy derivation of the large deviation Hamiltonian, and suggesting that this action should be valid way beyond the toy models considered by Rezakhanlou, b) the verification of the time-reversibility of the action and that the entropy is the quasipotential, c) the explanation of the interest of path large deviation to discuss the irreversibility paradox, d) stressing that a proper probabilistic interpretation of the molecular chaos hypothesis does not break time-reversal symmetry, e) the gradient structure of the Boltzmann equation, f) a unified view of known definitions and results about path large deviation theory: its relation with the structure of kinetic theories, entropy increase and decrease along relaxation and fluctuation paths, time-reversal symmetry of the action, relation with gradient structures, derivation of the action from the infinitesimal generator; which are essential to connect kinetic theories, large deviation theory, irreversibility and gradient flow structures.

Section \ref{sec:Introduction-to-Boltzmann's}
introduces the notations and is an introduction to the Boltzmann equation.
Section \ref{sec:Large deviations of small random perturbations}
is an introduction to the theory of path large deviations and the
related concepts: the quasipotential, the time reversibility of path
actions, the Hamilton-Jacobi equation, the monotonicity of quasipotential
evolutions, and so on. Section \ref{sec:The-reversible-Boltzmann}
derives the large deviation action for dilute gazes and studies its
main properties. Section \ref{sec:Gradient-structure} discusses
the gradient structure of the Boltzmann equation. 

\section{Introduction to the Boltzmann equation\label{sec:Introduction-to-Boltzmann's}}

This section introduces the notions of a dilute gas, collision rates,
diffusion cross section, the Boltzmann-Grad limit, which are key one's
in order to derive the Boltzmann equation. We give a heuristic derivation
of the Boltzmann equation in the spirit of Boltzmann's discussion.
Finally we present some of the key properties of the Boltzmann equation.

\subsection{Dilute gases \label{subsec:Dilute-gazes}}

\subsubsection{Orders of magnitudes and dimensionless numbers}

We consider the dynamics of a gas composed of atoms or molecules,
in the simplest possible framework. We neglect any internal degrees
of freedom. We assume that the dynamics is confined into a box of
volume $V$. We assume that the $N$ particles evolve through a Hamiltonian
dynamics with short range two body interactions, for instance hard
sphere collisions.

Several length scales are important to describe the gas: a typical
atom size $a$, that we will defined more precisely below in relation
with the diffusion cross section, a typical interparticle distance
$1/\rho^{1/3}$ where $\rho$ is the average gas density, the mean
free path $l$ which is the average length a particle travels between
two collisions, and a typical box size $V^{1/3}$. The gas is said
dilute if we have the following relation between those scales 
\[
a\ll\frac{1}{\rho^{1/3}}\ll l.
\]
We also assume that the box size is either of the order of the mean
free path $V^{1/3}\simeq l$ or much larger than the mean free path
$l\ll V^{1/3}$.

We note that those four length scales are not independent from each
other. If we consider particles in a box in the dilute gas limit,
we have four dimensional independent parameters: the volume $V$,
the particle number $N$ or equivalently the average density $\rho=N/V$,
the energy $E$ or equivalently the inverse temperature $\beta=1/k_{B}T$
where $k_{B}$ is Boltzmann's constant ($E=3N/2\beta$), and the typical
value of the cross section $a^{2}$. A typical velocity is $v_{T}=\sqrt{1/m\beta}$,
where $m$ is the particle mass. The mean free path $l$ can be determined
from these quantities. Its order of magnitude is obtained by considering
that the volume spanned be a particle between two collisions, $a^{2}l$,
should be equal to the typical volume occupied by each particle $1/\rho$.
The mean free path is thus $l=c/a^{2}\rho$, where $c$ is a non-dimensional
number that depends on the collision kernel, for instance $c=\sqrt{2}/8$
for hard sphere collisions. In the following, for simplicity we call
$l=1/a^{2}\rho$ the mean free path. We will also use a typical collision
time defined as $\tau_{c}=l/v_{T}=l\sqrt{m\beta}=\sqrt{m\beta}/a^{2}\rho$.
The typical values, for instance for hydrogen at the room temperature
and pressure are: $a\simeq1.4\,10^{-10}\mbox{m}$, $\rho\simeq2.\,10^{25}\,\text{m}^{-3}$
($1/\rho^{1/3}=3.7\,10^{-9}$), leading to $l=2.5\,10^{-6}\text{m}$
and volume $V=1\,\text{m}^{3}$. The typical particle velocity is
$v_{T}=1.6\,10^{3}\,\text{m.s}^{-1}$, which gives a collision time
of order of $\tau_{c}=6.7\,10^{-11}\text{s}$. 

We have four independent dimensional numbers with two sets of units.
There are thus two independent non dimensional numbers. We choose
$N$ as the first one. For the second one, we choose $\epsilon=1/l^{3}\rho=a^{2}/l^{2}=a^{6}\rho^{2}$
which is the order of magnitude of the inverse of the particle number
in a volume of size $l^{3}$. The limit $\epsilon\rightarrow0$ corresponds
to the condition $a\ll\frac{1}{\rho^{1/3}}\ll l.$ We note that the
$N\epsilon=V/l^{3}=\alpha^{-3}$, where $\alpha$ is called the Knudsen
number. We consider $\alpha$ fixed. Hence the constraint $V/l^{3}\simeq1$
(resp. $V/l^{3}\gg1$) is equivalent to $N\epsilon\simeq1$ or (resp.
$N\epsilon\gg1$). We call a limit as $\epsilon\rightarrow0$ and
$N\rightarrow\infty$ with either $N\epsilon=\alpha^{-3}$ fixed or
$N\epsilon\gg1$ a Boltzmann\textendash Grad limit. 

We note that in dimension $d$, the mean free path would be given
by $a^{d-1}l\rho=1$, and that the inverse of the number of particle
in a volume of linear size the mean free path would be $\epsilon=\left(\rho l^{d}\right)^{-1}=\left(a/l\right)^{d-1}=a^{d(d-1)}\rho^{d-1}$.
The Boltzmann-Grad limit would still be defined as $\epsilon\rightarrow0$
with $\epsilon N=\alpha^{-d}$ where $\alpha$ is a fixed constant,
or $\epsilon N\gg1$, in dimension $d$. We also note that it is customary,
in many mathematical papers and books, to define the Boltzmann-Grad
denoting $a=\epsilon'a_{0}$. In that case we would have $N\epsilon=N\epsilon'^{d-1}a_{0}^{d-1}/l^{d-1}$
and thus the Boltzmann-Grad limit is the limit $\epsilon'\rightarrow0$,
with $N\epsilon'^{d-1}=\alpha$ or $N\epsilon'^{d-1}\gg1$. For simplicity,
we do not follow those classical mathematics notations, because as
will be clear in the sequel, the large deviation rate will be of order
$1/\epsilon$ and the Gaussian fluctuations will be of order $\epsilon^{1/2}$,
whatever the dimension.

\subsubsection{Collision rate \label{subsec:Collisions}}

We consider a thread of particles with velocities $\mathbf{v}_{1}$
that meets a thread of particles with velocities $\mathbf{v}_{2}$.
We assume that particles of each velocity type are distributed according
to a homogeneous point Poisson process with densities $\rho(\mathbf{v}_{1})$
and $\rho(\mathbf{v}_{2})$ respectively. These particle distributions
will give rise to collisions where $(\mathbf{v}_{1},\mathbf{v}_{2})$
particle pairs undergo a random change towards pairs of the type $(\mathbf{v}'_{1},\mathbf{v}'_{2})$,
up to $(\mbox{d}\mathbf{v}'_{1},\mbox{d}\mathbf{v}'_{2})$. This occurs
at a rate per unit of time and unit of volume (in units $\mbox{m}^{-3}\mbox{s}^{-1}$)
which is proportional to the $\mathbf{v}_{1}$ incident particle density
$\rho(\mathbf{v}_{1})$, the $\mathbf{v}_{2}$ incident particle density
$\rho(\mathbf{v}_{2})$, $\mbox{d}\mathbf{v}'_{1},$ and $\mbox{d}\mathbf{v}'_{2}$.
The proportionality coefficient is called the collision kernel and
is denoted 
\begin{equation}
w_{0}\left(\mathbf{v}'_{1},\mathbf{v}'_{2};\mathbf{v}_{1},\mathbf{v}_{2}\right)/2.\label{eq:Microscopic_rate}
\end{equation}
As $w_{0}\left(\mathbf{v}'_{1},\mathbf{v}'_{2};\mathbf{v}_{1},\mathbf{v}_{2}\right)\mbox{d}\mathbf{v}'_{2}\mbox{d}\mathbf{v}'_{1}\rho(\mathbf{v}_{1})\rho(\mathbf{v}_{2})$
is in units $\mbox{m}^{-3}\mbox{s}^{-1}$, $w_{0}$ is in units $\mbox{m}^{-3}\mbox{s}^{5}$. 

The symmetry between particles 1 and 2 impose that 
\[
w_{0}(\mathbf{v}'_{1},\mathbf{v}'_{2};\mathbf{v}_{1},\mathbf{v}_{2})=w_{0}(\mathbf{v}'_{1},\mathbf{v}'_{2};\mathbf{v}_{2},\mathbf{v}_{1})=w_{0}(\mathbf{v}'_{2},\mathbf{v}'_{1};\mathbf{v}_{1},\mathbf{v}_{2}).
\]
The time reversal symmetry of the microscopic Hamiltonian dynamics
imposes that 
\[
w_{0}(\mathbf{v}'_{1},\mathbf{v}'_{2};\mathbf{v}_{1},\mathbf{v}_{2})=w_{0}(-\mathbf{v}{}_{1},-\mathbf{v}{}_{2};-\mathbf{v}'_{1},-\mathbf{v}'_{2}).
\]
The space rotation symmetry imposes that for any rotation $\mathbf{R}$
that belongs to the orthogonal group $SO(3)$
\[
w_{0}(\mathbf{v}'_{1},\mathbf{v}'_{2};\mathbf{v}_{1},\mathbf{v}_{2})=w_{0}(\mathbf{R}\mathbf{v}{}_{1},\mathbf{R}\mathbf{v}{}_{2};\mathbf{R}\mathbf{v}'_{1},\mathbf{R}\mathbf{v}'_{2}).
\]
The combination of the time reversal symmetry and of the space rotation
symmetry for $\mathbf{R}=-\mathbf{I}$, where $\mathbf{I}$ is the
identity operator, implies the inversion symmetry 
\begin{equation}
w_{0}(\mathbf{v}'_{1},\mathbf{v}'_{2};\mathbf{v}_{1},\mathbf{v}_{2})=w_{0}(\mathbf{v}{}_{1},\mathbf{v}{}_{2};\mathbf{v}'_{1},\mathbf{v}'_{2}).\label{eq:inversion}
\end{equation}
The local conservation of momentum and energy implies that
\begin{equation}
w_{0}(\mathbf{v}'_{1},\mathbf{v}'_{2};\mathbf{v}_{1},\mathbf{v}_{2})=\sigma(\mathbf{v}'_{1},\mathbf{v}'_{2};\mathbf{v}_{1},\mathbf{v}_{2})\delta\left(\mathbf{v}_{1}+\mathbf{v}_{2}-\mathbf{v}'_{1}-\mathbf{v}'_{2}\right)\delta\left(\mathbf{v}_{1}^{2}+\mathbf{v}_{2}^{2}-\mathbf{v'}_{1}^{2}-\mathbf{v'}{}_{2}^{2}\right),\label{eq:cross_section}
\end{equation}
where $\sigma$ is the diffusion cross section. $\sigma$ is of the
order of $a^{2}$ where $a$ is a typical atom size. Integrating this
expression over $\mathbf{v}'_{1}$ and $\mathbf{v}'_{2}$, using that
$\sigma$ is of the order of $a^{2}$, we see that the average number
of collisions that a single particle pair with velocity $\left(\mathbf{v}_{1},\mathbf{v}_{2}\right)$
undergoes per unit of time is of order $a^{2}\left\Vert \mathbf{v}_{1}-\mathbf{v}_{2}\right\Vert /V$.

\subsection{Distribution functions \label{subsec:Distribution-function}}

We consider $N$ particles. Each particle $1\leq n\leq N$ has a position
$\mathbf{r}_{n}(t)$ and a velocity $\mathbf{v}_{n}(t)$, with initial
conditions $\mathbf{r}_{n}(0)$ and $\mathbf{v}_{n}(0)$. We define
the empirical distribution as 
\begin{equation}
f_{e}(\mathbf{r},\mathbf{v},t)\equiv\sum_{n=1}^{N}\delta\left(\mathbf{v}-\mathbf{v}_{n}(t)\right)\delta\left(\mathbf{r}-\mathbf{r}_{n}(t)\right).\label{eq:Empirical_Velocity_Distribution-1-1}
\end{equation}
The normalization is such that $\int\mbox{d}\mathbf{v}\mbox{d}\mathbf{r}\,f_{e}(\mathbf{v})=N$.
The number of particles with position $\mathbf{r}_{1}$ up to $\mbox{d}\mathbf{r}_{1}$
and velocity $\mathbf{v}_{1}$ up to $\mbox{d}\mathbf{v}_{1}$ is
$f_{e}\left(\mathbf{r}_{1},\mathbf{v}_{1}\right)\mbox{d}\mathbf{v}{}_{1}\mbox{d}\mathbf{r}_{1}$. 

We assume that the particles evolve according to their own velocity
and their mutual collision only. The evolution of the empirical density
is given by 
\[
\frac{\partial f_{e}}{\partial t}+\mathbf{v}.\frac{\partial f_{e}}{\partial\mathbf{r}}=\mathcal{C}
\]
where $\mathcal{C}$ accounts for the collision effects.\\

We consider an ensemble of initial conditions $\left(\mathbf{r}_{n}(0),\mathbf{v}_{n}(0)\right)_{1\leq n\leq N}$,
distributed according to a measure\\
$f_{N}\left(\mbox{\ensuremath{\mathbf{r}}}_{1},\mbox{\ensuremath{\mathbf{v}}}_{1},...,\mbox{\ensuremath{\mathbf{r}}}_{N},\mbox{\ensuremath{\mathbf{v}}}_{N},t=0\right)\prod_{n}\mbox{d}\mathbf{r}_{n}\mbox{d}\mathbf{v}_{n}$.
We assume that with respect to this measure, the probability to have
the empirical distribution equal to a distribution $f(\mathbf{r},\mathbf{v})$
concentrates close $f_{0}(\mathbf{r}_{0},\mathbf{v}_{0})$. We do
not give a precise definition of this concentration property here,
however one can think of a large deviation principle similar to the
one that will be used latter on in section \ref{sec:The-reversible-Boltzmann}.
We denote this concentration property $f_{e}\succ f_{0}$. $f_{0}(\mathbf{r},\mathbf{v})$
can be alternatively understood as the average of $f_{e}$ when averaging
over the initial conditions described by the density $f_{N}$: $f_{0}=\mathbb{E}_{i}(f_{e})$.

It is natural to expect this concentration property to hold for later
times $t>0$, if $t$ is not too large, in the Boltzmann-Grad limit.
We then expect $f_{e}(t)\succ f(t)$, or alternatively $f(t)=\mathbb{E}_{i}(f_{e}(t))$.
The aim of Boltzmann equation is to describe the temporal evolution
of $f$ in the Boltzmann-Grad limit.

\subsection{Boltzmann's equation \label{subsec:Intro-Boltzmann's-equation}}

\subsubsection{Main assumptions and heuristic derivation\label{subsec:Main-assumptions-Boltzmann}}

We recall that we consider an ensemble of initial conditions such
that the empirical distribution concentrates for time $t=0$: $f_{e}(t=0)\succ f_{0}$.
We are looking for the equation of the distribution function $f$
such that $f_{e}(t)\succ f(t)$.

In order to derive Boltzmann's equation, we will make the following
four assumptions, which are a reformulation of Boltzmann's initial
hypotheses.

i) The collision duration can be neglected compared to the average
collision time. The geometry of the collisions is also neglected (point
particle assumption).

ii) The probability of three particle encounters is extremely low
and will be neglected.

iii) \textbf{\emph{Molecular chaos hypothesis:}} In classical physics
textbooks in kinetic theory, the Boltzmann equation is often obtained
formally through the BBGKY hierarchy. Then the molecular chaos hypothesis
is usually stated as the property that the two point correlation function
$f_{2}$ can be approximated by the product of two one point correlations
functions $f$. This amounts to a property about the statistical independence
of the colliding particles. This assumption allows one to close the
BBGKY hierarchy and to obtain the Boltzmann equation. Mathematical
proofs of the validity of the Boltzmann equation seeks to justify
this assumption. For the sake of the following discussion, we will
not go through the BBGKY hierarchy; we then define the molecular chaos
hypothesis directly as a property of the statistics of the particle
collisions. We will use the following assumption, that we still call
\textbf{\emph{the molecular chaos hypothesis}}: at any time, for an
overwhelming number of initial conditions, the effect of the collisions
of type $\left(\mathbf{v}_{1},\mathbf{v}_{2}\right)\rightarrow\left(\mathbf{v}'_{1},\mathbf{v}'_{2}\right)$
on the distribution $f_{e}$ can be quantified as if, locally in position
space, the particles with velocity $\mathbf{v}_{1}$ up to $\mbox{d}\mathbf{v}_{1}$
and $\mathbf{v}_{2}$ up to $\mbox{d}\mathbf{v}_{2}$ would be statistically
mutually independent and each distributed according to a local Poisson
point process in position space with density $\rho(\mathbf{v}_{1})=f(\mathbf{r},\mathbf{v}_{1},t)\mbox{d}\mathbf{v}_{1}$
and $\rho(\mathbf{v}_{2})=f(\mathbf{r},\mathbf{v}_{2},t)\mbox{d}\mathbf{v}_{2}$
respectively. Then the definition of the collision kernel is relevant
and the collision rate $w$ is the only relevant physical quantity.
This hypothesis iii) is the counterpart of the molecular chaos hypothesis
(stosszahlansatz) used by Boltzmann. It is a very natural hypothesis that the local Poisson statistics should be created dynamically by streaming through particle velocity. This should also be the case for a collisionless ideal gas, as first suggested by Bogolyubov \cite{bogoliubov1962problems}. A dynamical foundation for collision less ideal gases has been given by Eyink and Spohn \cite{eyink2000space}.

iv) \textbf{Law of large numbers:} In this section, following Boltzmann,
we compute only the average number of collisions and we do not consider
the possible fluctuations of the collision number around this average
number.\\

With these hypothesis, we can write the effect of collisions. The
distribution function $f(\mathbf{r},\mathbf{v},t)$ changes both because
particles of velocity $\mathbf{v}$ are created or disappear through
collisions. Using also the inversion symmetry (\ref{eq:inversion}),
we readily obtain the Boltzmann equation 
\[
\frac{\partial f}{\partial t}+\mathbf{v}.\frac{\partial f}{\partial\mathbf{r}}=\int\mbox{d}\mathbf{v}{}_{2}\mbox{d}\mathbf{v}'_{1}\mbox{d}\mathbf{v}'_{2}\,w_{0}(\mathbf{v}'_{1},\mathbf{v}'_{2};\mathbf{v},\mathbf{v}_{2})\left[f\left(\mathbf{\mathbf{v}'}_{1},\mathbf{r}\right)f\left(\mathbf{v}'_{2},\mathbf{r}\right)-f\left(\mathbf{\mathbf{v}},\mathbf{r}\right)f\left(\mathbf{v}_{2},\mathbf{r}\right)\right].
\]

The key hypothesis in this heuristic derivation of Boltzmann's equation
are the molecular chaos hypothesis iii) and law of large numbers iv).
While Boltzmann's molecular chaos hypothesis is usually the subject
of much attention, hypothesis iv), is usually not commented upon. \textbf{Following
assumption iv), only the average effect of collisions is taken into
account in the evolution of the empirical distribution function. This
is a natural hypothesis as in the Boltzmann-Grad limit the number
of collisions that occur on a mesoscopic volume with a mean free path
size is extremely large. Hypothesis iv) can thus be interpreted as
a law of large numbers. While natural, this law of large numbers assumption
necessarily implies that Boltzmann's equation can describe at best
only the average or typical evolution of the empirical distribution. }

\textbf{As discussed in section \ref{sec:The-reversible-Boltzmann},
hypothesis iv) can be relaxed, in order to analyze the stochastic
process of the evolution of the distribution $f$ that results from
the hypothesis i), ii) and iii) only.}

\subsubsection{Main properties of the Boltzmann equation\label{subsec:Properties-of-Boltzmann's-Eq}}

We list the main properties of the Boltzmann equation

i) The total mass $\int\mbox{d}\mathbf{v}\mbox{d}\mathbf{r}\,f$ is
conserved. As a consequence of the local conservation laws encoded
in Eq. (\ref{eq:cross_section}), the total momentum $\mathbf{P}=\int\mbox{d}\mathbf{v}\mbox{d}\mathbf{r}\,\mathbf{v}f$
and the total kinetic energy $E=\int\mbox{d}\mathbf{v}\mbox{d}\mathbf{r}\,f\mathbf{v}^{2}/2$
are locally conserved. 

ii) The entropy 
\[
S[f]=-\int\mbox{d}\mathbf{v}\mbox{d}\mathbf{r}\,f\log(f)
\]
increases: $\frac{\mbox{d}S}{\mbox{d}t}\geq0$.

iv) Maxwell-Boltzmann distributions $f_{MB}\left(\mathbf{r},\mathbf{v}\right)=A\exp\left[-\beta\frac{\left(\mathbf{v}-\mathbf{U}\right)^{2}}{2}\right]$
are stationary solutions of the Boltzmann equation.

iii) With some more assumptions on the collision kernel, one can prove
that the entropy is strictly increasing except for Maxwell-Boltzmann
distributions. In those cases, the Boltzmann equation then converges
towards the Maxwell-Boltzmann distribution for which the mass, energy,
and momentum are equal to the initial ones.\\

For completeness, we should discuss boundary conditions. In the following
we will discuss the cases either of an infinite box, finite box with
elastic collisions of the particles at the box wall, or particles
on a three dimensional torus (periodic boundary conditions). Then,
while the mass and the energy will be globally conserved, the momentum
may or may not be globally conserved, depending on the boundary conditions.
We do not discuss further the boundary conditions in the following.

\section{Large deviations produced by a large number of small amplitude independent
moves \label{sec:Large deviations of small random perturbations}}

When the evolution of a stochastic process is the consequence of the
effect of a large number of small amplitude statistically independent
moves, in the limit of a large number of moves, a law of large numbers
naturally follows. It is often very important to understand the large
deviations with respect to this law of large number. For continuous
time Markov processes, for instance diffusions with small noises,
or more generally locally infinitely divisible processes, a general
framework can be developed in order to estimate the probability of
large deviations. In this section, we present this framework briefly,
mainly the hypothesis on the generator that leads to large deviations
and the formula for the large deviation action that quantifies the
probability of those large deviations. We then define and recall the
main properties of action minima, relaxation and fluctuation paths,
quasipotentials, relation between action symmetry and detailed balance,
and conservation laws, which are crucial for the analysis of large
deviation properties. While the simple and synthetic presentation
of this section, containing the key ideas without the mathematical
discussion is original, most of the material in this section is classical.
Our main reference is Freidlin\textendash Wentzell textbook \cite{FW2012}. The point of view we proposed based on the Hamiltonian (\ref{eq:H-Generator}) has been much developed by  \cite{feng2006large}. Many aspects, like the relation between the action symmetry and detailed
balance, are not treated in \cite{FW2012} or \cite{feng2006large}. Those aspects are most
probably classical too, some but not all of them are discussed in \cite{bertini2015macroscopic}, section II.C. The following discussion is kept at a formal level, avoiding
any mathematical technicalities that may hide the most important ideas.

\subsection{Large deviation rate functions from the infinitesimal generator of
a continuous time Markov process \label{sec:Large_Deviations_Generator-1}}

We consider $\left\{ X_{\epsilon}(t)\right\} _{0\leq t\leq T}$, a
family of continuous time Markov processes parametrized by a real
number $\epsilon$. For instance $X_{\epsilon}(t)\in\mathbb{R}^{n}$.
$\mathbb{E}_{x}$ denotes the conditional average over the Markov
process given that $X_{\epsilon}(0)=x$. We denote $G_{\epsilon}$
the infinitesimal generator of the process (the definition of the
infinitesimal generator of a continuous time Markov process is given
in section \ref{subsec:Infinitesimal-generator-of}). We assume that
for all $p\in\mathbb{R}^{n}$ the limit 
\begin{equation}
H(x,p)=\lim_{\epsilon\downarrow0}\epsilon G_{\epsilon}\left[\mbox{e}^{\frac{p.x}{\epsilon}}\right]\mbox{e}^{-\frac{p.x}{\epsilon}}\label{eq:H-Generator}
\end{equation}
exists. Then the family $X_{\epsilon}$ verifies a large deviation
principle with rate $\epsilon$ and rate function 
\begin{equation}
L\left(x,\dot{x}\right)=\sup_{p}\left\{ p\dot{x}-H(x,p)\right\} .\label{eq:Lagrangian}
\end{equation}
This means that the probability that the path $\left\{ X_{\epsilon}(t)\right\} _{0\leq t<T}$
is in a neighborhood of $\left\{ X(t)\right\} _{0\leq t<T}$ verifies
\begin{equation}
P\left[\left\{ X_{\epsilon}(t)\right\} _{0\leq t<T}=\left\{ X(t)\right\} _{0\leq t<T}\right]\underset{\epsilon\downarrow0}{\asymp}\exp\left(-\frac{\int_{0}^{T}\mbox{d}t\,L\left(X,\dot{X}\right)}{\epsilon}\right),\label{eq:Large deviations path}
\end{equation}
where the symbol $\underset{\epsilon\downarrow0}{\asymp}$ is a logarithm
equivalence ($f_{\epsilon}\underset{\epsilon\downarrow0}{\asymp}\exp(g/\epsilon)\iff\lim_{\epsilon\downarrow0}\epsilon\log f_{\epsilon}=g$).

This result is proven for specific cases (diffusions, locally infinitely
divisible processes) in the Theorem 2.1, page 127, of the third edition
of Freidlin-Wentzell textbook \cite{FW2012}. A heuristic derivation
is given in section \ref{sec:Large_Deviations_Generator}, page \pageref{sec:Large_Deviations_Generator},
of this paper. The examples of a diffusion and of a Poisson process
are discussed in section \ref{sec:Examples}. We apply this framework
to the fluctuations of the empirical distribution of a dilute gas
in section \ref{sec:The-reversible-Boltzmann}.

In formula (\ref{eq:H-Generator}) the infinitesimal generator is
tested through the function $\mbox{e}^{\frac{p.x}{\epsilon}}$ . In
the small $\epsilon$ limit, this tests those changes of the observable
which are of order of $\epsilon$. The $\epsilon$ prefactor in the
right hand side of equation (\ref{eq:H-Generator}) means that the
overall effect of these small changes of order $\epsilon$ is expected
to be of order $1/\epsilon$. $H$ in formula (\ref{eq:H-Generator})
thus accounts for the effects of a large number (of order $1/\epsilon$)
of small amplitude statistically independent moves (each one of order
$\epsilon$). 

\subsection{An example: large deviation for radioactive decay\label{subsec:An-example:-large}}

We now consider the example of the large deviations of the radioactive
decay of an ensemble of $N$ independent particles. This example will
be very useful in the following because of its analogy with collisions
in dilute gas, to be studied in section \ref{sec:The-reversible-Boltzmann}.

\subsubsection{Definition of a radioactive decay}

We consider a radioactive decay from a state $x=1$ to a state $x=0$
at rate $\lambda$. By radioactive decay, we mean a pure death-process,
where the event $1\rightarrow0$ occurs at a rate $\lambda$. Given
that $x(t=0)=1$, the distribution of times at which the particle
will decay is $\lambda\exp(-\lambda\tau)$. Let us assume that we
have $N$ particles $x_{n}$, with $1\leq n\leq N$, each of which
undergoes independently a radioactive decay at rate $\lambda$. We
consider the ratio of particles that have not yet decayed at time
$t,$
\begin{equation}
X_{N}(t)=\frac{1}{N}\sum_{n=1}^{N}x_{n}(t).\label{eq:X_N}
\end{equation}
By the law of large numbers, we immediately know that $X_{N}(t)$ converge
for large $N$ to the average $\bar{X}(t)$ that satisfies 
\begin{equation}
\frac{\mbox{d}\bar{X}}{\mbox{d}t}=-\lambda\bar{X}\label{eq:Exponential_Decay}
\end{equation}
and, using $\bar{X}(0)=1$, we obtain $\bar{X}(t)=\exp\left(-\lambda t\right)$.

What is the probability to observe a path for $X_{N}$ that is different
from $\bar{X}$? We see that any particle decay changes the value
of $X_{N}$ by a factor $1/N$. A change of order $1$ of the variable
$X_{N}$ is thus the result of $N$ independent events, each one of
amplitude $1/N$. It is thus natural to expect a large deviation estimate
that states that 
\[
P\left[\left\{ X_{N}(t)\right\} _{0\leq t\leq T}=\left\{ X(t)\right\} \right]\underset{N\uparrow\infty}{\asymp}\exp\left(-NI(X)\right).
\]
How to compute $I(X)$? For this problem, because the $N$ particles
are statistically independent, the answer can be obtained directly
in many different ways. In the next section, we use the framework
of section \ref{sec:Large_Deviations_Generator-1} in order to compute
$I$.

\subsubsection{Large deviation action for the radioactive decay of $N$ particles}

As the $N$ particles are independent, the variable $X_{N}$ (\ref{eq:X_N})
defines a continuous time Markov process. From the definition of the
radioactive decay, and the definition of the infinitesimal generator
of a Markov process (see section \ref{subsec:Infinitesimal-generator-of}),
we can compute straightforwardly (see section \ref{subsec:Infinitesimal-generator-of})
the infinitesimal generator for the evolution of the variable $X_{N}$:
\[
G_{N}\left[\phi\right](x)=N\lambda x\left[\phi\left(x-\frac{1}{N}\right)-\phi\left(x\right)\right],
\]
where $x=n/N$ with $n$ any integer number with $1\leq n\leq N$,
and $\phi$ is a real valued function on $\left[0,1\right]$. We also
have $G_{N}\left[\phi\right](0)=0$.

We get
\[
\lim_{N\rightarrow\infty}\frac{1}{N}G_{N}\left[\mbox{e}^{Npx}\right]\mbox{e}^{-Npx}=H(x,p)\equiv\lambda x\left(\mbox{e}^{-p}-1\right).
\]
From the general discussion of section \ref{sec:Large_Deviations_Generator-1},
we thus conclude that $I\left[X\right]=\int_{0}^{T}\mbox{d}t\,L\left(X,\dot{X}\right)$,
where $L$ is the Legendre\textendash Fenchel transform of $H$. We
obtain 
\[
L(x,\dot{x})=\dot{x}+\lambda x-\dot{x}\log\left(-\frac{\dot{x}}{\lambda x}\right)\,\,\,\mbox{if}\,\,\,\dot{x}<0\,\,\,\mbox{and}\,\,\,-\infty\,\,\,\mbox{otherwise}.
\]
We note that the large deviation rate function has the property that
$X_{N}$ is necessarily decreasing, as imposed by its definition (\ref{eq:X_N}).

As explained in the following sections, the most probable evolution
(the law or large numbers) verifies the relaxation path equation 
\[
\dot{x}=R(x)=\frac{\partial H}{\partial p}(x,0)=-\lambda x,
\]
which is actually the same as (\ref{eq:Exponential_Decay}), as expected.

\subsection{Quasipotential, Hamilton Jacobi equation, time reversal symmetry, relaxation and fluctuation
paths\label{subsec:Quasipotential,-Hamilton-Jacobi}}

We consider the properties of a stochastic process for which rare fluctuations
are described, at the level of large deviations, by the action 
\begin{equation}
\mathcal{A}\left[X\right]=\int_{0}^{T}\mbox{d}t\,L\left(X,\dot{X}\right)=\sup_{P}\int_{0}^{T}\mbox{d}t\,\left[P\dot{X}-H\left(X,P\right)\right].\label{eq:Action}
\end{equation}
The most probable evolution corresponding to the action (\ref{eq:Action}),
and with initial condition $X_{r}(t=0)=x$ is called a \textbf{\emph{relaxation
path issued from $x$}}. It solves $\dot{X_{r}}=R(X_{r})$, with initial
condition $X_{r}(0,x)=x$, where $R(x)=\arg\inf_{\dot{x}}L\left(x,\dot{x}\right)$. 

We assume that the stochastic process $X_{\varepsilon}$ has a stationary
distribution $P_{s,\epsilon}$ which dynamics follows the large deviation
principle
\begin{equation}
P_{s,\epsilon}(x)\equiv\mathbb{E}\left[\delta\left(X_{\epsilon}-x\right)\right]\underset{\epsilon\downarrow0}{\asymp}\exp\left(-\frac{U(x)}{\epsilon}\right),\label{eq:Quasipotential_Definition-1}
\end{equation}
where $U$ is called the \textbf{\emph{quasipotential}}. In order
to simplify the following discussion, we also assume that the relaxation
equation has a single fixed point $x_{0}$ and that any solution to
the relaxation equation converges to $x_{0}$. This assumption can
be easily relaxed as done for instance in \cite{FW2012}. However
this assumption is actually true for the dilute gas dynamics and simplifies
much of the discussion. Then the quasipotential verifies 
\[
U(x)=\inf_{\left\{ \left\{ X(t)\right\} {}_{-\infty\leq t\leq0}\left|X(-\infty)=x_{0}\,\,\,\mbox{and}\,\,\,X(0)=x\right.\right\} }\int_{-\infty}^{0}\mbox{d}t\,L\left(X,\dot{X}\right).
\]
The minimizer of this variational problem, that is the most probable
path starting from $x_{0}$ and ending at $x$, is denoted $X_{f}(t,x)$
and is called the \textbf{\emph{fluctuation path ending at $x$.}}\\

We have the following properties which are direct consequences of
the definitions of $H$ and $L$, and whose proofs are given in sections
\ref{subsec:Quasipotential,-relaxation-paths} to \ref{subsec:A-sufficient-condition}:
\begin{enumerate}
\item $H$ is a convex function of the variable $p$ and $H(x,0)=0$, see
sec. \ref{subsec:Some-properties-of}.
\item $L\geq0$, see sec. \ref{subsec:Some-properties-of}.
\item For any $x,$ $\dot{x}$ and $p$
\begin{equation}
p\dot{x}\leq L(x,\dot{x})+H(x,p),\label{eq:Convexity_Inequality}
\end{equation}
see sec. \ref{subsec:Some-properties-of}.
\item The relaxation paths solve the equation $\dot{x}=R(x)$ with $\inf_{\dot{x}}L(x,\dot{x})=0=L\left(x,R(x)\right)$,
and $R(x)=\frac{\partial H}{\partial p}(x,0)$, see sec. \ref{subsec:Relaxation-paths}. 
\item The quasipotential solves the stationary \textbf{Hamilton\textendash Jacobi
equation} 
\begin{equation}
H(x,\nabla U)=0,\label{eq:Stationary_Hamilton_Jacobi}
\end{equation}
see sec. \ref{subsec:Quasipotential}.
\item \textbf{The fluctuation paths} solve the first order equation 
\[
\dot{X_{f}}=F(X_{f})\equiv\frac{\partial H}{\partial p}\left(X_{f},\nabla U(X_{f})\right),
\]
see sec. \ref{subsec:Fluctuation-paths}.
\item As $H$ is convex, the quasipotential decreases along the relaxation
paths 
\[
\frac{\mbox{d}U}{\mbox{d}t}(X_{r})=H(X_{r},0)-H(X_{r},\nabla U(X_{r}))+\frac{\partial H}{\partial p}\left(X_{r},0\right).\nabla U(X_{r})\leq0,
\]
see sec. \ref{subsec:Decay/increase-of-the}.
\item As $H$ is convex, the quasipotential increases along the fluctuation
paths
\[
\frac{\mbox{d}U}{\mbox{d}t}(X_{f})=H(X_{f},0)-H(X_{f},\nabla U(X_{f}))+\frac{\partial H}{\partial p}\left(X_{f},\nabla U(X_{f})\right).\nabla U(X_{f})\geq0,
\]
see sec. \ref{subsec:Decay/increase-of-the}.
\item \textbf{\emph{Time reversal symmetry and detailed balance }}(see sec. \ref{subsec:Detailed-balance}).
A stationary continuous time Markov process is said to be time reversible if its backward and forward histories have the same probabilities. This property is equivalent to the detailed balance condition (see sec. \ref{subsec:Detailed-balance}), or to the fact that the infinitesimal generator of the time reversed process is identical to the infinitesimal generator of the initial process.  In sec. \ref{subsec:Detailed-balance}, we prove that at the level of large deviations, the time reversal symmetry (or detailed balance condition) reads either 
\[
\mbox{for any \ensuremath{x} and \ensuremath{\dot{x}}, }L(x,\dot{x})-L(x,-\dot{x})=\dot{x}.\nabla U,
\]
or equivalently 
\begin{equation}
\mbox{for any \ensuremath{x} and p, }H\left(x,-p\right)=H\left(x,p+\nabla U\right).\label{eq:Bilan_Detaille}
\end{equation}
\item \textbf{\emph{Generalized detailed balance }}(see sec. \ref{subsec:Generalized-detailed-balance}).
For most physical systems the notion of time reversibility has to be extended, for instance in order to take into account that the velocity sign has to be changed in systems with inertia, or other fields have to be modified in the time-reversal symmetry. This is true for the time-reversal symmetry of dynamical systems, for instance of mechanical systems described by Hamiltonian equations, but also for the time-reversal symmetry of Markov processes. Such a generalized definition of time reversal symmetry is classical both in the physics and the mathematics literature, see for instance \cite{Gardiner_1994_Book_Stochastic}.

Let $I$ be the involution that characterizes the time-reversal symmetry
(for instance the map that correspond to velocity or momentum inversion
in many systems). We assume that $I$ is self adjoint for the scalar
product, that is $I(x).p=I(p).x$. A continuous time Markov process is said to be time-reversal symmetric in the generalized sense if its backward histories with the application of $I$ and its forward histories have the same probabilities. This definition is equivalent to either a generalized detailed balance condition that involves $I$ (see sec. \ref{subsec:Generalized-detailed-balance}) or to the fact that the infinitesimal generator of the time reversed process is identical to the generator of the initial process up to application of the involution $I$. As explained in sec. \ref{subsec:Generalized-detailed-balance}, the detailed balance conditions for the quasipotential $U$ combined with the involution $I$ are
$U(x)=U\left(I\left[x\right]\right)$ on one hand and 
\[
L(x,\dot{x})-L(x,-I\left[\dot{x}\right])=I\left[\dot{x}\right].\nabla U
\]
or equivalently 
\[
H\left(I\left[x\right],-I\left[p\right]\right)=H\left(x,p+\nabla U\right),
\]
on the other hand.
\item As can be easily checked, if either the detailed balance or the generalized
detailed balance conditions are verified, then $U$ satisfies the
stationary Hamilton-Jacobi equation (\ref{eq:Stationary_Hamilton_Jacobi}).
\item If the detailed balance condition is verified, and if $U$ is the
quasipotential, then for a path $\left\{ X(t)\right\} _{0\leq t\leq T}$
and its time reversed one $\left\{ I\left[X(T-t)\right]\right\} _{0\leq t\leq T}$
we have the symmetry for the path probability 
\[
P\left[X_{\epsilon}(t)=\left\{ X(t)\right\} _{0\leq t\leq T}\right]\mbox{e}^{-\frac{U(X(0))}{\epsilon}}=P\left[X_{\epsilon}(t)=\left\{ I\left[X(T-t)\right]\right\} _{0\leq t\leq T}\right]\mbox{e}^{-\frac{U(I\left[X(T)\right])}{\epsilon}},
\]
see sec. \ref{subsec:Detailed-balance-general}.
\item \textbf{\emph{Conserved quantities }}(see sec. \ref{subsec:Conservation-law-Hamiltonien}\textbf{\emph{)}}\emph{.}
At the level of the large deviations, the condition for $C(x)$ to
be a conserved quantity is either 
\[
\mbox{for any}\,\,x\,\,\mbox{and}\,\,p,\,\,\,L(x,\dot{x})=+\infty\,\,\,\mbox{if}\,\,\,\dot{x}.\nabla C(x)\neq0,
\]
or 
\begin{equation}
\mbox{for any}\,\,x\,\,\mbox{and}\,\,p,\,\,\,\frac{\partial H}{\partial p}\left(x,p\right).\nabla C=0.\label{eq:Conservation_Law_H-1}
\end{equation}
\item \textbf{\emph{A sufficient condition for $U$ to be the quasipotential
}}(see sec. \ref{subsec:A-sufficient-condition}). If $U$ solves
the Hamilton\textendash Jacobi equation, if $U$ has a single minimum
$x_{0}$ with $U(x_{0})=0$, and if for any $x$ the solution of the
reverse fluctuation path dynamics $\dot{X}=-F(X)=-\frac{\partial H}{\partial p}\left(X,\nabla U(X)\right)$
with $X(0)=x$ converges to $x_{0}$ for large times, then $U$ is
the quasipotential.
\end{enumerate}

\section{Large deviations for dilute gas dynamics and the Boltzmann equation\label{sec:The-reversible-Boltzmann}}

In section \ref{sec:Introduction-to-Boltzmann's} we gave a heuristic
derivation of the Boltzmann equation, following the hypotheses described
in section \ref{subsec:Main-assumptions-Boltzmann}, page \pageref{subsec:Main-assumptions-Boltzmann}.
For this derivation, the key hypotheses are: the molecular chaos hypothesis
iii), and the law of large numbers iv) assumptions (please see section
section \ref{subsec:Main-assumptions-Boltzmann}, page \pageref{subsec:Main-assumptions-Boltzmann}).
In this section we relax the law of large number assumption iv), but
still assume the molecular chaos hypothesis. This allows us to compute
the action that describes the large deviations of the empirical distribution,
under the molecular chaos hypothesis assumption. 

We also discuss the properties of the large deviation action, prove
its time-reversal symmetry, explain why the entropy is the negative
of the quasipotential, and discuss the physical consequences in relation
with the irreversibility paradox.

\subsection{Derivation of the action for dilute gas dynamics}

In the following we work with non dimensional variables and functions.
A natural time scale for the dynamics is the typical collision time
$\tau_{c}$, while a natural spatial scale is the mean free path $l$
; we thus take as a time unit $\tau_{c}$ and space unit $l$ such
that with these units a typical velocity is of order one. As in section
\ref{sec:Introduction-to-Boltzmann's}, we consider $N$ particles
in a volume $V$. Each particle $1\leq n\leq N$ has a position $\mathbf{r}_{n}(t)$
and a velocity $\mathbf{v}_{n}(t)$. We assume that these $N$ particles
have a total mass $M_{0}$, momentum $\mathbf{P}_{0}$ and kinetic
energy $E_{0}$ ; those values are conserved by the dynamics.

In order to clearly identify the large deviation rate, we want to
make sure that all quantities remain of order 1 in the Boltzmann\textendash Grad
limit $\epsilon\rightarrow0$ and $N\rightarrow\infty$ with $\epsilon N=\alpha^{-3}$
(or $\epsilon N\gg1$). As $\rho l^{3}=1/\epsilon$, typical densities
are of order $1/\epsilon$, and typical velocities are of order one
for our choice of units. If we would define the empirical distribution
as in equation (\ref{eq:Empirical_Velocity_Distribution-1-1}) it
would be of order $1/\epsilon$. We thus define a rescaled non dimensional
empirical distribution as 
\begin{equation}
f_{\epsilon}(\mathbf{r},\mathbf{v},t)\equiv\epsilon\sum_{n=1}^{N}\delta\left(\mathbf{v}-\mathbf{v}_{n}(t)\right)\delta\left(\mathbf{r}-\mathbf{r}_{n}(t)\right).\label{eq:Empirical_Velocity_Distribution-1}
\end{equation}
The normalization is now such that $\int\mbox{d}\mathbf{v}\mbox{d}\mathbf{r}f_{\epsilon}(\mathbf{v})=N\epsilon=V/l^{3}$,
such that $f_{\epsilon}$ remains finite in the limit $\epsilon\rightarrow0$.
Similarly, as the cross section is of order $a^{2}$, as velocities
are of order one, the collision kernel is order $a^{2}/l^{2}=\epsilon$.
We thus work with a rescaled collision kernel $w$ defined such that
a couple of particles with velocities $\left(\mathbf{v}_{1},\mathbf{v}_{2}\right)$
in a volume element $\mbox{d}\mathbf{r}$ has a microscopic collision
of the type $\left(\mathbf{v}_{1},\mathbf{v}_{2}\right)\rightarrow\left(\mathbf{v}'_{1},\mathbf{v}'_{2}\right)$
that occurs with a rate per unit of volume equal to 
\begin{equation}
\epsilon w\left(\mathbf{v}'_{1},\mathbf{v}'_{2};\mathbf{v}_{1},\mathbf{v}_{2}\right)\mbox{d}\mathbf{v}'_{1}\mbox{d}\mathbf{v}'_{2}.\label{eq:Microscopic_rate-1-2}
\end{equation}
This new definition has to be compared with equation (\ref{eq:Microscopic_rate})
($w_{0}=\epsilon w$). As $f_{\epsilon}=\epsilon f_{e}$, and $w=w_{0}/\epsilon$
where $f_{e}$ and $w_{0}$ are the section \ref{sec:Introduction-to-Boltzmann's}
distribution function and collision kernel respectively, we note that
Boltzmann's equation is unchanged by this rescaling.

This natural rescaling makes clear that the large deviation rate will
be $\epsilon$. Indeed, equations (\ref{eq:Empirical_Velocity_Distribution-1})
and (\ref{eq:Microscopic_rate-1-2}) readily show that individual
collisions for each particle occur at a small rate of order $\epsilon$
and produce a small effect on the distribution function $f$ of order
of $\epsilon$. As a consequence, on a time scale of order 1, $1/\epsilon$
collisions occur each producing a change of order $\epsilon$ of the
empirical distribution function, and thus producing an overall change
of the distribution function $f_{\epsilon}$ of order $1$. This clearly
corresponds to a large deviation scaling with rate $\epsilon$.\\

We now proceed with the derivation of the action. We compute, at a
large deviation level, the probability that the empirical density
$f_{\epsilon}$ remains close to a distribution function $f$. According
to the molecular chaos hypothesis iii) (please see section
\ref{subsec:Main-assumptions-Boltzmann}, page \pageref{subsec:Main-assumptions-Boltzmann}),
in order to evaluate the collision rates, we identify $f_{\epsilon}$
with $f$ and make the hypothesis of a local point Poisson process.
The number of particles with position $\mathbf{r}_{1}$ up to $\mbox{d}\mathbf{r}_{1}$
and velocity $\mathbf{v}_{1}$ up to $\mbox{d}\mathbf{v}_{1}$ is
$f\left(\mathbf{r}_{1},\mathbf{v}_{1}\right)\mbox{d}\mathbf{v}{}_{1}\mbox{d}\mathbf{r}_{1}/\epsilon$.
The number of particle couples, per unit volume, with the two particles
with position $\mathbf{r}$ up to $\mbox{d}\mathbf{r}$, and with
the first particle with velocity $\mathbf{v}_{1}$ up to $\mbox{d}\mathbf{v}_{1}$
and the second particle with velocity $\mathbf{v}_{2}$ up to $\mbox{d}\mathbf{v}_{2}$
is $f(\mathbf{r},\mathbf{\mathbf{v}_{1})}f\left(\mathbf{r},\mathbf{v}_{2}\right)\mbox{d}\mathbf{v}{}_{1}\mbox{d}\mathbf{v}{}_{2}\mbox{d}\mathbf{r}/2\epsilon^{2}$.
Combining this with (\ref{eq:Microscopic_rate-1-2}) we obtain that
the collision rate of the type $\left(\mathbf{v}_{1},\mathbf{v}_{2}\right)\rightarrow\left(\mathbf{v}'_{1},\mathbf{v}'_{2}\right)$
in the volume element $\mbox{d}\mathbf{r}$ centered at a point $\mathbf{r}$
is 

\[
\frac{w(\mathbf{v}'_{1},\mathbf{v}'_{2};\mathbf{v}_{1},\mathbf{v}_{2})}{2\epsilon}f(\mathbf{r},\mathbf{\mathbf{v}_{1})}f\left(\mathbf{r},\mathbf{v}_{2}\right)\mbox{d}\mathbf{v}{}_{1}\mbox{d}\mathbf{v}{}_{2}\mbox{d}\mathbf{v}'_{1}\mbox{d}\mathbf{v}'_{2}\mbox{d}\mathbf{r}.
\]

Each individual collision $\left(\mathbf{v}_{1},\mathbf{v}_{2}\right)\rightarrow\left(\mathbf{v}'_{1},\mathbf{v}'_{2}\right)$
in the volume element $\mbox{d}\mathbf{r}$ centered at a point $\mathbf{r}$
changes the empirical velocity distribution (\ref{eq:Empirical_Velocity_Distribution-1})
from $f(\mathbf{.},\mathbf{.)}$ to 
\[
f(\mathbf{.},\mathbf{.)}-\epsilon\left[\delta\left(\mathbf{.}-\mathbf{v}_{1}\right)\delta\left(\mathbf{.}-\mathbf{r}\right)-\delta\left(\mathbf{.}-\mathbf{v}_{2}\right)\delta\left(\mathbf{.}-\mathbf{r}\right)\right.
\]
\[
\left. +\delta\left(\mathbf{.}-\mathbf{v}'_{2}\right)\delta\left(\mathbf{.}-\mathbf{r}\right)+\delta\left(\mathbf{.}-\mathbf{v'}_{1}\right)\delta\left(\mathbf{.}-\mathbf{r}\right)\right].
\]
 The infinitesimal generator of the empirical distribution is thus
\[
G\left[\phi\right]\left[f\right]=G_{R}\left[\phi\right]\left[f\right]+G_{C}\left[\phi\right]\left[f\right]
\]
with
\begin{multline}
G_{R}\left[\phi\right]\left[f\right]=\frac{1}{2\epsilon}\int\mbox{d}\mathbf{v}{}_{1}\mbox{d}\mathbf{v}{}_{2}\mbox{d}\mathbf{v}'_{1}\mbox{d}\mathbf{v}'_{2}\mbox{d}\mathbf{r}\,w(\mathbf{v}'_{1},\mathbf{v}'_{2};\mathbf{v}_{1},\mathbf{v}_{2}) f(\mathbf{r},\mathbf{\mathbf{v}_{1})}f\left(\mathbf{r},\mathbf{v}_{2}\right)\\
\times\left\{ \phi\left[f(\mathbf{.)}+\epsilon\left[-\delta\left(\mathbf{.}-\mathbf{v}_{1}\right)\delta\left(\mathbf{.}-\mathbf{r}\right)-\delta\left(\mathbf{.}-\mathbf{v}_{2}\right)\delta\left(\mathbf{.}-\mathbf{r}\right) \right. \right. \right.\\
\left. \left.\left. + \delta\left(\mathbf{.}-\mathbf{v}'_{2}\right)\delta\left(\mathbf{.}-\mathbf{r}\right)+\delta\left(\mathbf{.}-\mathbf{v'}_{1}\right)\delta\left(\mathbf{.}-\mathbf{r}\right)\right]\right]-\phi\left[f\right]\right\} ,
\end{multline}
and 
\begin{equation}
G_{C}\left[\phi\right]\left[f\right]=-\int\mbox{d}\mathbf{r}\mbox{d}\mathbf{v}\,\mathbf{v}.\frac{\partial f}{\partial\mathbf{r}}(\mathbf{r},\mathbf{v})\frac{\delta\phi}{\delta f(\mathbf{r},\mathbf{v})},\label{eq:Inhomogeneous_Boltzmann_Generator}
\end{equation}
where $G_{R}$ is the collision contribution to the generator, and
$G_{C}$ accounts for the transport by the ballistic motion of the
particles and the associated free transport of the distribution. Section
\ref{subsec:The-infinitesimal-generator} justifies that $G_{C}$
takes this form for a deterministic free transport.

In view of the derivation of the large deviation action from the infinitesimal
generator, discussed in section \ref{sec:Large_Deviations_Generator-1},
more precisely equation (\ref{eq:H-Generator}), we define 
\begin{equation}
H\left[f,p\right]\equiv\epsilon G\left[\mbox{e}^{\frac{1}{\epsilon}\int\mathrm{d}\mathbf{r}\mathrm{d}\mathbf{\mathbf{v}}p(\mathbf{r},\mathbf{v})f(\mathbf{r},\mathbf{v})}\right]\mbox{e}^{-\frac{1}{\epsilon}\int\mathrm{d}\mathbf{r}\mathrm{d}\mathbf{\mathbf{v}}p(\mathbf{r},\mathbf{v})f(\mathbf{r},\mathbf{v})},\label{eq:Hamiltonien_Inhomogeneous_Definition}
\end{equation}
where $p$ is the momentum conjugated to $f$. Hence 
\begin{equation}
H\left[f,p\right]=H_{C}\left[f,p\right]+H_{R}\left[f,p\right]\label{eq:Hamiltonian_Inhomogeneous}
\end{equation}
with 
\begin{multline}
H_{R}\left[f,p\right]=\frac{1}{2}\int\mbox{d}\mathbf{v}{}_{1}\mbox{d}\mathbf{v}{}_{2}\mbox{d}\mathbf{v}'_{1}\mbox{d}\mathbf{v}'_{2}\mbox{d}\mathbf{r}\,w(\mathbf{v}'_{1},\mathbf{v}'_{2};\mathbf{v}_{1},\mathbf{v}_{2})\\
\times f(\mathbf{r},\mathbf{\mathbf{v}_{1})}f\left(\mathbf{r},\mathbf{v}_{2}\right)\left\{ \mbox{e}^{\left[-p\left(\mathbf{r},\mathbf{v}_{1}\right)-p\left(\mathbf{r},\mathbf{v}_{2}\right)+p\left(\mathbf{r},\mathbf{v}'_{1}\right)+p\left(\mathbf{r},\mathbf{v}'_{2}\right)\right]}-1\right\} \label{eq:Hamiltonian-Collisions}
\end{multline}
and 
\begin{equation}
H_{C}\left[f,p\right]=-\int\mbox{d}\mathbf{r}\mbox{d}\mathbf{v}\,p(\mathbf{r},\mathbf{v})\mathbf{v}.\frac{\partial f}{\partial\mathbf{r}}(\mathbf{r},\mathbf{v}),\label{eq:Hamiltonian-Transport}
\end{equation}
where the subscript $R$ in $H_{R}$ means reversible and the subscript
$C$ in $H_{C}$ means conservative (see below).

From the discussion of section \ref{sec:Large_Deviations_Generator-1},
we thus conclude that we have a large deviation principle with rate
$\epsilon$ and action 
\begin{equation}
\mathcal{A}\left[f\right]=\int_{0}^{T}\mbox{d}t\,L\left[f,\dot{f}\right]=\int_{0}^{T}\mbox{d}t\,\sup_{p}\left[\int p\left(\mathbf{r},\mathbf{v}\right)\dot{f}\left(\mathbf{r},\mathbf{v}\right)\,\mbox{d}\mathbf{r}\mbox{d}\mathbf{v}-H\left[f,p\right]\right].\label{eq:Action_Inhomogeneous_Boltzmann}
\end{equation}
We have thus computed the action for the evolution of the empirical
distribution of a dilute gas, assuming the molecular chaos hypothesis. 

\subsection{Time reversal symmetry, quasipotential, fluctuation and relaxation
paths\label{Time_reversal_symmetry_Boltzmann}}

Let us discuss this action properties. First, the corresponding most
probable evolution is given by $\frac{\partial f}{\partial t}=\frac{\delta H}{\delta p}\left[f,0\right]$
(see point 4 in section \ref{subsec:Quasipotential,-Hamilton-Jacobi},
or section \ref{subsec:Relaxation-paths}). It is easily checked that
this most probable evolution is Boltzmann's equation 
\[
\frac{\partial f}{\partial t}+\mathbf{v}.\frac{\partial f}{\partial\mathbf{r}}=\int\mbox{d}\mathbf{v}{}_{2}\mbox{d}\mathbf{v}'_{1}\mbox{d}\mathbf{v}'_{2}\,w(\mathbf{v}'_{1},\mathbf{v}'_{2};\mathbf{v},\mathbf{v}_{2})\left[f\left(\mathbf{\mathbf{v}'}_{1},\mathbf{r}\right)f\left(\mathbf{v}'_{2},\mathbf{r}\right)-f\left(\mathbf{\mathbf{v}},\mathbf{r}\right)f\left(\mathbf{v}_{2},\mathbf{r}\right)\right],
\]
as expected.

We define the specific entropy in the dilute gas limit as 
\begin{equation}
S\left[f\left|f_{0}\right.\right]=-\int\mbox{d}\mathbf{r}\mbox{d}\mathbf{v}\,f\log\left(\frac{f}{f_{0}}\right),\label{eq:Entropy}
\end{equation}
where $f_{0}=f_{M_{0},\mathbf{P}_{\text{0}},E_{0}}$ is the Maxwell-Boltzmann
distribution with mass $M_{0}$, momentum $\mathbf{P}_{0}$ and kinetic
energy $E_{0}$. $S$ is a relative entropy of $f$ with respect to
$f_{0}$. It is also the Boltzmann $H$ function up to an additive
constant.

We now discuss the time reversal symmetry. We define the time reversal
involution (the velocity inversion involution) $I$ by $I\left[f\right](\mathbf{r},\mathbf{v})=f(\mathbf{r},-\mathbf{v})$.
Then clearly $I^{2}=Id$ and $I$ is self adjoint for the $L_{2}$
scalar product. It is easily checked that $H_{R}\left(I\left[f\right],-I\left[p\right]\right)=H_{R}\left(f,-p\right)=H_{R}\left(f,p-\frac{\delta S}{\delta f}\right)$,
and that $H_{C}\left(I\left[f\right],-I\left[p\right]\right)=H_{C}\left(f,p-\frac{\delta S}{\delta f}\right)$,
where the last equality is a consequence of the conservation of the
entropy by the free transport operator. As a consequence 
\begin{equation}
H\left(I\left[f\right],-I\left[p\right]\right)=H\left(f,p-\frac{\delta S}{\delta f}\right),\label{eq:H_Boltzmann_time_reversal_symmetry}
\end{equation}
which is the detailed balance condition with $-S$ as the quasipotential
(see points 9) and 10) in section \ref{subsec:Quasipotential,-Hamilton-Jacobi}). We have thus checked that the large-deviation action is time-reversal symmetric. 
We note that the collision term and the transport term act differently
with respect to the time reversal symmetry. The collision term is
both time reversal symmetric and time reversal symmetric with respect
to the involution $I$, while the transport term is time reversal
symmetric only with respect to the involution $I$. 

Using point 11) of section \ref{subsec:Quasipotential,-Hamilton-Jacobi},
we thus conclude that $-S$ solves the Hamilton\textendash Jacobi
equation.

We now discuss conserved quantities. From point 13) of section \ref{subsec:Quasipotential,-Hamilton-Jacobi},
a functional $C\left[f\right]$ is conserved for the large deviation
action (\ref{eq:Action_Inhomogeneous_Boltzmann}) if and only if $\int\mbox{d}\mathbf{r}\mbox{d}\mathbf{v}\,\frac{\delta H}{\delta p}\frac{\delta C}{\delta f}=0$.
This is easily checked for the conservation laws for the mass 
\[
M=\int\mbox{d}\mathbf{r}\mbox{d}\mathbf{v}\,f,
\]
the momentum 
\[
\mathbf{P}=\int\mbox{d}\mathbf{r}\mbox{d}\mathbf{v}\,\mathbf{v}f,
\]
and the kinetic energy 
\[
E=\frac{1}{2}\int\mbox{d}\mathbf{r}\mbox{d}\mathbf{v}\,\mathbf{v}^{2}f,
\]
as consequences of the microscopic conservation laws encoded in (\ref{eq:cross_section}).
We denote $M_{0}$, $\mathbf{P}_{0}$, and $E_{0}$ the initial values
of the mass, momentum, and kinetic energy respectively (we note that
depending on the boundary conditions, $\mathbf{P}$ may not be a conserved
quantity, we let the reader adapt the discussion to this case).

Given that $-S$ solves Hamilton\textendash Jacobi's equation and
given the conservation laws, it is natural to assume that 
\[
U\left[f\right]=\left\{ \begin{array}{l}
-S\left[f\left|f_{0}\right.\right]\,\,\,\text{if}\,\,\,M\left[f\right]=M_{0},\,\,\,\mathbf{P}\left[f\right]=\mathbf{P}_{0},\,\,\,\mbox{and}\,\,\,E\left[f\right]=E_{0}\\
-\infty\,\,\,\mbox{otherwise}.
\end{array}\right.
\]
is the quasipotential. This is also what should be expected from equilibrium
statistical mechanics, in the microcanonical ensemble. We note that
due to the convexity of $U$, we are in the case when $U$ has a single
minimum. This minimum of $U$ is $f_{0}=f_{M_{0},\mathbf{P}_{\text{0}},E_{0}}$,
the Maxwell-Boltzmann distribution with mass $M_{0}$, momentum $\mathbf{P}_{0}$
and kinetic energy $E_{0}$.

We want to use point 14) of section \ref{subsec:Quasipotential,-Hamilton-Jacobi}
to establish that $U$ is indeed the quasipotential. We note that
due to the generalized time reversal symmetry of the action with respect
to the entropy (\ref{eq:H_Boltzmann_time_reversal_symmetry}), in
order to verify the hypothesis of point 14), it is sufficient to verify
that the solution to the Boltzmann equation starting from any distribution
function $f$ with mass mass $M_{0}$, momentum $\mathbf{P}_{0}$
and kinetic energy $E_{0}$ converges for large time to the Maxwell-Boltzmann
distribution $f_{M_{0},\mathbf{P}_{\text{0}},E_{0}}$. This property
is actually true for generic non-degenerate kernels $w$. For a more
precise discussion and hypothesis and the discussion of the rate of
relaxation to equilibrium, please see \cite{desvillettes2005trend}
and \cite{lu2015measure}, and references therein.

We thus conclude that in generic cases, as soon as the solution to
Boltzmann's equation starting from any distribution function $f$
with mass mass $M_{0}$, momentum $\mathbf{P}_{0}$ and kinetic energy
$E_{0}$ converges for large time to the Maxwell-Boltzmann distribution
$f_{M_{0},\mathbf{P}_{\text{0}},E_{0}}$, $U$ is the quasipotential.\\

Finally, from the general discussion of section \ref{subsec:Quasipotential,-Hamilton-Jacobi},
points 7), 8) and 12), we deduce that the entropy increases along
the relaxation paths (solution of Boltzmann's equation), that the
fluctuation paths are the time reversed relaxation paths using the
involution $I$, and thus that the entropy decreases along the fluctuation
paths.

\subsection{Physical discussion}

\subsubsection{The irreversibility paradox}

Our main result is that the probability that $\left\{ f_{\epsilon}(t)\right\} _{0\leq t\leq T}$
remains in a neighborhood of a prescribed path $\left\{ f(t)\right\} _{0\leq t\leq T}$
verifies 
\begin{equation}
\mathbb{P}\left[\left\{ f_{\epsilon}(t)\right\} _{0\leq t<T}=\left\{ f(t)\right\} _{0\leq t<T}\right]\underset{\epsilon\downarrow0}{\asymp}\exp\left(-\frac{\int_{0}^{T}\mbox{d}t\,L\left[f,\dot{f}\right]}{\epsilon}\right),\label{eq:Large_Deviation_Action-1}
\end{equation}
where the large deviation Lagrangian $L$ is the Legendre-Fenchel
transform of the Hamiltonian $H$, and $H$ is defined through (\ref{eq:Hamiltonian_Inhomogeneous}-\ref{eq:Action_Inhomogeneous_Boltzmann}).
This defines a statistical field theory whose large deviations are
given through $H$.

We have proven in the previous section that:

i) The stochastic process with action (\ref{eq:Large_Deviation_Action-1})
is time reversible. This is a consequence of the time reversal symmetry
of $L$ or $H$. Any path $\left\{ f_{\epsilon}(t)\right\} _{0\leq t<T}$
is possible with a probability given by (\ref{eq:Large_Deviation_Action-1}).

ii) The most probable evolution from any state $f_{0}$ (a relaxation
path) is the solution of the Boltzmann equation with initial condition
$f_{0}$.

iii) The entropy $S[f]=-\int\mbox{d}\mathbf{v}\mbox{d}\mathbf{r}\,f\log(f)$
is the negative of the quasipotential for the statistical field theory.
As a consequence it increases monotonically along any relaxation path,
and thus along the evolution through the Boltzmann equation.

We believe that those properties clarify in a definitive way the seemingly
paradox of irreversibility. It is fully compatible with the classical
understanding following Boltzmann original ideas. We can summarize
this result by the following sentences. The macrostates evolve according
to a completely reversible stochastic process. Then the irreversibility
is not a consequence of looking at the system at a macroscopic scale.
However the change of scale for the description of the physical phenomena,
produces entropic factors such that some evolutions of the macroscopic
variables appear as much more probable than others. The irreversibility
is thus a consequence of two combined effect: looking at macroscopic
scale only and looking at only the most probable evolution. It is
not a consequence of the change of scale for the description of the
physical phenomena by itself. 

We stress also that the molecular chaos hypothesis used to derive
the large deviation action (\ref{eq:Large_Deviation_Action-1}) does
not break the time reversal symmetry.

\subsubsection{The relation with fluctuating hydrodynamics and macroscopic fluctuation
theory}

Our work has been initially inspired by the works of Derrida and Lebowitz
on the large deviation for stochastic particle dynamics, for instance
for the Asymmetric Exclusion Process \cite{Derrida_Lebowitz_1998_PhRvL},
and the works of the Rome group on macroscopic fluctuation theory
\cite{bertini2015macroscopic}. One can consider the result (\ref{eq:Large_Deviation_Action-1})
and (\ref{eq:Hamiltonian_Inhomogeneous}-\ref{eq:Action_Inhomogeneous_Boltzmann})
as a macroscopic fluctuation theory for the dynamics of dilute gases.
Our is a macroscopic fluctuation theory which is relevant for a large
class of physical systems which are dilute in the sense that the physical
constituents interact in a way a dilute gas does.

The large deviation principle (\ref{eq:Large_Deviation_Action-1})
can also be considered as a starting point for deriving large deviation
results at a different scale. For instance can we deduce Landau's
fluctuating hydrodynamics \cite{Landau_Lifshitz_1996_Book} and the
related large deviations through a coarse-graining of the action (\ref{eq:Large_Deviation_Action-1})?
Can we deduce the large deviation for a piston separating two boxes
with dilutes gazes through a coarse-graining of (\ref{eq:Large_Deviation_Action-1})?
Those are natural questions that will be addressed in future works.

\subsubsection{Validity of the molecular chaos hypothesis}

The main assumption to derive the large deviation action for dilute
gas is the molecular chaos hypothesis. It is obviously outside of
the scope of this paper to justify it. Let us however give few arguments
to support it. 

1) \textbf{The molecular chaos hypothesis is the most natural one.}
First it has to be stressed that the molecular chaos hypothesis is
the most natural order zero assumption for dilute gas dynamics, for
both the law of large numbers and large deviation estimates. Most
of the time, particles that undergo collisions come from spatial region
very far apart and will rarely collide again. The correlation are
therefore expected to be extremely weak in the dilute gas limit. We
have to notice however, that making the molecular chaos hypothesis
in order to study large deviations is a stronger assumption than making
the assumption for the law of large number only. We believe that in
the same way Boltzmann's equation is believed to actually describe
most physical situations for dilute gas, the large deviation action
(\ref{eq:Hamiltonian_Inhomogeneous}-\ref{eq:Action_Inhomogeneous_Boltzmann})
should describe large deviations in most physical situations. We discuss
few possible exceptions like for instance shocks in the following.

Up to now, it has proven extremely difficult to quantify the effects
of weak correlations and some possible corrections to Boltzmann's
equation. It will probably be at least as difficult to discuss the
related questions for the large deviation rate function.

2)\textbf{ Perfect gas entropy, quasipotential and the Boltzmann-Grad
limit.} Independently of the dynamics, the specific entropy for a
perfect gas (\ref{eq:Entropy}) is the leading order Boltzmann entropy
of the macrostate $f$ for the microcanonical measure, in the Boltzmann-Grad
limit $\epsilon\downarrow0$ with $\epsilon N\sim1$ or $\epsilon N\gg1$.
This was probably first noticed by Boltzmann himself, and this fact
is usually explained in basic statistical mechanics lecture following
the counting arguments of Boltzmann. For this reason, it was natural
to expect this expression for the entropy to be the negative of the
quasipotential for the large deviation rate function, up to conservation
laws. The fact that the large deviation action (\ref{eq:Hamiltonian_Inhomogeneous}-\ref{eq:Action_Inhomogeneous_Boltzmann})
has the perfect gas entropy (\ref{eq:Entropy}) as a quasipotential
stress consistency of the large deviation rate function with its expected
properties. 

3) \textbf{Smoothness of $f$ and the molecular chaos hypothesis.
}The molecular chaos hypothesis iii), in section \ref{subsec:Intro-Boltzmann's-equation}
on page \pageref{subsec:Intro-Boltzmann's-equation}, assumes that
the effect of collisions can be well approximated as if the distribution
of particles with velocity $\mathbf{v}$ would be locally the one
of a Poisson point process with density derived from $f$. For the
Boltzmann equation, this hypothesis would most probably not remain
consistent if $f$ would not be smooth. Similarly, for the large deviation action where $f$ is now
prescribed a-priori, this hypothesis requires to consider classes
of smooth enough functions $f$ for consistency. Given the dynamical
mechanisms at hand (free transport and collisions), it seems natural
to assume that $f$ has variations that do not extend below the mean
free path scale, this assumption being consistent with the molecular
chaos hypothesis. We are unfortunately not able to be more precise
on that point. 

\subsubsection{Gaussian fluctuations}

The Gaussian fluctuations around the solution of the Boltzmann equation
have been studied by a number of authors. For instance close to equilibrium
they are described in Spohn's book \cite{Spohn_1991} or have been
studied mathematically \cite{bodineau2017hard}. Far for equilibrium
macrostates $f$, the stochastic process of Gaussian fluctuations
close to solutions of the Boltzmann equation has been described by
physicists, please see for instance \cite{gaspard2013time} and references
therein. For instance equation (61) in section 4 of \cite{gaspard2013time}
gives a formula for the two point correlation function of the Gaussian
noise needed to obtain the fluctuating Boltzmann equation that describe
those Gaussian fluctuations. 

An important remark is that our large deviation principle (\ref{eq:Large_Deviation_Action-1})
is fully compatible with the previous description of those Gaussian
fluctuations, as far as small deviation are concerned. What is meant
here is that by linearization of the Hamiltonian (\ref{eq:Hamiltonian_Inhomogeneous}-\ref{eq:Action_Inhomogeneous_Boltzmann}),
one recovers exactly the two point correlations described by equation
(61) of \cite{gaspard2013time}. I do not reproduce the related easy
computations here.

\subsubsection{Large deviation principle for the Kac's model }

Kac's model \cite{kac1956foundations,mischler2013kac} is a continuous
time Markov chain that mimics the homogeneous Boltzmann's equation,
where only the velocity variable are represented. We consider $N$
particles with velocities $(\mathbf{v}_{1},...,\mathbf{v}_{N})\in\mathbb{R}^{dN}$.
The dynamics of the Kac's model is defined by the infinitesimal generator
\begin{multline*}
G_{N}\left[\phi_{N}\right](\mathbf{v}_{1},...,\mathbf{v}_{N})=\frac{1}{2N}\sum_{n,m=1,n\neq m}^{N}\int\mbox{d}\mathbf{v}'_{1}\mbox{d}\mathbf{v}'_{2}\,w(\mathbf{v}'_{1},\mathbf{v}'_{2};\mathbf{v}_{n},\mathbf{v}_{m})\\
\times \left\{ \phi_{N}\left[C_{nm,\mathbf{v'}_{1}\mathbf{v'}_{2}}(\mathbf{v}_{1},...,\mathbf{v}_{N})\right]-\phi_{N}(\mathbf{v}_{1},...,\mathbf{v}_{N})\right\} ,
\end{multline*}
where $\phi_{N}:\mathbb{R}^{dN}\rightarrow\mathbb{R}$ is a test function
in the $N$ velocity space, and $C_{nm,\mathbf{v'}_{1}\mathbf{v'}_{2}}$
is the operator that replace the velocities $\mathbf{v}_{n}$ and
$\mathbf{v}_{m}$ by $\mathbf{v}'_{1}$ and $\mathbf{v}'_{2}$ respectively
($C_{nm}$ is defined for $n\neq m$ ; if $n<m$, $C_{nm,\mathbf{v'}_{1}\mathbf{v'}_{2}}(\mathbf{v}_{1},...,\mathbf{v}_{N})=(\mathbf{v}_{1},...,\mathbf{v}_{n-1},\mathbf{v'}_{1},\mathbf{v}_{n+1},...,\mathbf{v}_{m-1},\mathbf{v}'_{2},\mathbf{v}_{m+1},...,\mathbf{v}_{N})$).
This is the generator of a dynamics for which a collision of the type
$\left(\mathbf{v}_{1},\mathbf{v}_{2}\right)\rightarrow\left(\mathbf{v}'_{1},\mathbf{v'}_{2}\right)$
occurs with rate $w(\mathbf{v}'_{1},\mathbf{v}'_{2};\mathbf{v}_{n},\mathbf{v}_{m})/2N$.

By an easy computation analogous to the one discussed in section \ref{sec:The-reversible-Boltzmann},
one easily justify formally that the empirical density $f_{e}(\mathbf{v},t)\equiv\frac{1}{N}\sum_{n=1}^{N}\delta\left(\mathbf{v}-\mathbf{v}_{n}(t)\right)$
verifies a large deviation principle 
\begin{equation}
\mathbb{P}\left[\left\{ f_{\epsilon}(t)\right\} _{0\leq t<T}=\left\{ f(t)\right\} _{0\leq t<T}\right]\underset{\epsilon\downarrow0}{\asymp}\exp\left(-\frac{\int_{0}^{T}\mbox{d}t\,L\left[f,\dot{f}\right]}{\epsilon}\right),\label{eq:Large_Deviation_Action-1-1}
\end{equation}
where the large deviation Lagrangian $L$ is the Legendre-Fenchel
transform of the Hamiltonian $H$, and $H$ is defined by 
\begin{multline*}
H\left[f,p\right]=\frac{1}{2}\int\mbox{d}\mathbf{v}{}_{1}\mbox{d}\mathbf{v}{}_{2}\mbox{d}\mathbf{v}'_{1}\mbox{d}\mathbf{v}'_{2}w(\mathbf{v}'_{1},\mathbf{v}'_{2},\mathbf{v}_{1},\mathbf{v}_{2})\\
\times f\left(\mathbf{v}_{1}\right)f\left(\mathbf{v}_{2}\right)\left[\mbox{e}^{p\left(v'_{1}\right)+p\left(v'_{2}\right)-p\left(v_{1}\right)-p\left(v_{2}\right)}-1\right].
\end{multline*}

This Hamiltonian is also, obviously, the one obtained from the Hamiltonian
of the dilute gas large deviations (\ref{eq:Hamiltonian_Inhomogeneous}-\ref{eq:Action_Inhomogeneous_Boltzmann}),
when restricting evolutions to spatially homogeneous solutions.

\section{Gradient structure for the Boltzmann equation\label{sec:Gradient-structure}}

There exists a close relation between the gradient structure of a
many classical PDEs and the large deviations for the empirical density
of related $N$ particle dynamics. These PDE are obtained as laws
of large number for the particle dynamics. When one describes the
fluctuations of the empirical density at the level of large deviations,
one obtain a large deviation rate function. The PDE appears as the
most probable evolution, the action minimum. The large deviation rate
function that measures the probability to depart from the most probable
evolution also provides a natural metric structure for the PDE. When
the large deviation action is time reversible, one can prove that
the PDE solution is a gradient flow, with an energy (energy from the
point of view of gradient flows) which is the quasipotential, and
with a norm (in the quadratic case), or a metric structure (in the
general case), that derives from the large deviation rate functions.
These provide systematic connections between the limit of particle
systems to PDE, large deviation theory, and gradient structure.

The gradient structure of an equation may be extremely useful both
physically and mathematically: the energy (or quasipotential) landscape
gives a first qualitative idea of the dynamics, the gradient structure
explains convergence properties and sometimes convergence speeds.
At a mathematical level, the gradient structure may be used to prove
existence results in a very natural way.

Such gradient flow structures had been observed independently of large
deviation theory, for instance by Otto \cite{otto2001geometry}. For
instance the heat equation, appears as a gradient structure with the
entropy as a quasipotential and the Wasserstein distance as the metric
structure, or the Vlasov\textendash Mac-Kean equation has a gradient
structure with the free energy as the quasipotential and the
Wasserstein distance for the metric structure (see for instance \cite{villani2008optimal}).
Other examples include the Allen\textendash Cahn equation, equations
for porous media, and so on. 

All those structures can be obtained (and thus explained) from the
large deviation principle for the evolution of the empirical density.
Some aspects of this connection were first understood for the case
of Vlasov\textendash Mac-Kean equation (Brownian particles with mean
field interactions) in works by Dawson and Gärtner \cite{Dawson}.
The connection between large deviations and gradient flows has been
explained very clearly, with mathematical rigor in some specific cases,
by Mielke, Peletier and Renger \cite{mielke2014relation}. We give
the definition of gradient flows in section \ref{subsec:Gradient-flow}
and briefly explain the relation between time reversible large deviation
principles for paths and gradient flows in section \ref{subsec:Gradient-flow-for}.

As a consequence of the general relation between large deviation for
paths and gradient structures, using the large deviation principle
for the empirical density for dilute gazes, we can infer that a gradient
structure exists for the Boltzmann equation. The Boltzmann equation
is not time reversible, and the corresponding large deviation principle
verifies a generalized detailed balance. As a consequence, the Boltzmann
equation is not strictly speaking a gradient flow, but the collision
operator of the Boltzmann equation has a gradient structure: the collision
part is a gradient of the entropy with respect to a metric structure.
As explained in section \ref{subsec:Gradient-structure-for-Boltzmann},
this metric structure is not simple however. Moreover, the homogeneous
Boltzmann equation is a gradient flow. This is explained in section
\ref{subsec:Gradient-structure-for-Boltzmann}.

\subsection{Gradient flows\label{subsec:Gradient-flow}}

In this section we define gradient flows. We consider $\psi^{*}$
a function of two variables $x$ and $p$, and assume that $\psi^{*}$
is a convex function with respect to the second variable. We call
$\psi$ the convex conjugated of $\psi^{*}$ with respect to the second
variable 
\begin{equation}
\psi(x,v)=\sup_{p}\left\{ vp-\psi^{*}(x,p)\right\} .\label{eq:Legendre_Fenchel_psi}
\end{equation}
$\psi^{*}$ is called the dissipation function. $\psi^{*}$ is assumed
to be non negative and such that $\psi^{*}(x,0)=0$. $\psi^{*}(x,0)=0$
implies that $\psi$ is also non negative, and $\psi^{*}\geq0$ implies
that $\psi(x,0)=0$. 

A gradient flow with energy $E$ and dissipation function $\psi^{*}$
is defined as a solution $\left\{ x(t)\right\} _{0\leq t\leq T}$
to the equation 
\begin{equation}
E(x(T))-E(x(0))+\int_{0}^{T}\mbox{dt}\,\left[\psi(x,\dot{x})+\psi^{*}(x,-\nabla E)\right]=0,\label{eq:Gradient_Flow}
\end{equation}
 where $\dot{x}$ is the time derivative of $x$ and $\nabla$ the
gradient for the canonical norm. The non negativity of $\psi$ and
$\psi^{*}$ insures that the energy decreases with time.

When time derivation of (\ref{eq:Gradient_Flow}) can be justified,
we obtain 
\[
\dot{x}.\nabla E+\psi(x,\dot{x})+\psi^{*}(x,-\nabla E)=0.
\]
Using Fenchel's inequality, we obtain that
\[
\dot{x}.\nabla E+\psi(x,\dot{x})+\psi^{*}(x,-\nabla E)\geq0
\]
is always verified. When $\psi$ is differentiable, the equality is
verified whenever
\[
\dot{x}=\frac{\partial\psi^{*}}{\partial p}\left(x,-\nabla E\right).
\]
This is the gradient flow differential equation and (\ref{eq:Gradient_Flow})
is a weak form of this differential equation.

A special case of interest is when $\psi^{*}$ is quadratic in the
second variable, for instance $\psi^{*}(x,p)=pAp$ where $A$ is a
linear operator acting on the $p$ space. Then we obtain the classical
gradient flow 
\[
\dot{x}=-2A\nabla E,
\]
with respect to the norm given by $\psi^{*}(x,p)=pAp$.

\subsection{Gradient flows for time reversible path large deviations\label{subsec:Gradient-flow-for}}

Let assume a large deviation rate function for path $\left\{ X_{\epsilon}(t)\right\} $,
with Hamiltonian $H(x,p)$, and its associated Lagrangian $L(x,\dot{x})$
obtained by Legendre\textendash Fenchel transform: 
\begin{equation}
P\left[\left\{ X_{\epsilon}(t)\right\} _{0\leq t<T}=\left\{ X(t)\right\} _{0\leq t<T}\right]\underset{\epsilon\downarrow0}{\asymp}\exp\left(-\frac{\int_{0}^{T}\mbox{d}t\,L\left(X,\dot{X}\right)}{\epsilon}\right),\label{eq:Large deviations path-1}
\end{equation}
with $L\left(x,\dot{x}\right)=\sup_{p}\left\{ p\dot{x}-H(x,p)\right\} $,
and with quasipotential $U$ such that $H(x,\nabla U)=0$. We assume
the time reversal symmetry with respect to the quasipotential $U$
(see Eq. (\ref{eq:Bilan_Detaille}) page \pageref{eq:Bilan_Detaille}):
for any $p$, $H\left(x,p+\nabla U\right)=H\left(x,-p\right)$. Equivalently
$H$ is a symmetric function of the second variable with respect to
$\nabla U/2$: for any $p$, $H\left(x,\nabla U/2+p\right)=H\left(x,\nabla U/2-p\right)$.
Using that $H(x,0)=0$ (point 1 of page \pageref{subsec:Quasipotential,-Hamilton-Jacobi}),
we note that $U$ solves the stationary Hamilton\textendash Jacobi
equation $H\left(x,\nabla U\right)=0$.

In this section we justify that the relaxation paths for those dynamical
large deviations are gradient flows for the dissipation function $\psi^{*}$
given by (\pageref{eq:psi_star_H}) and the energy $E=U/2$.\\

We define the dissipation function $\psi^{*}$ by 
\begin{equation}
\psi^{*}\left(x,p\right)=H\left(x,\frac{\nabla U}{2}+p\right)-H\left(x,\frac{\nabla U}{2}\right)\label{eq:psi_star_H}
\end{equation}
The convexity of $\psi^{*}$ follows from the convexity of $H$. We
have $\psi^{*}\left(x,0\right)=0$. Using the convexity of $H$, we
have that $H(x,\nabla U/2)\leq \frac{1}{2}\left[H(x,\nabla U/2+p)+H(x,\nabla U/2-p)\right]=H(x,\nabla U/2+p)$
where the last equality is a consequence of the symmetry of $H$.
This insures that $\psi^{*}$ is non negative. Hence $\psi^{*}$ has
all the property of a dissipation function. We note moreover that
$\psi^{*}$ is even. 

From the definition of $\psi$ as the Legendre-Fenchel transform of
$\psi^{*}$, $L$ as the Legendre Fenchel transform of $H$, from
(\ref{eq:psi_star_H}), we get $\psi(x,\dot{x})=L(x,\dot{x})-\dot{x}.\nabla U/2+H\left(x,\nabla U/2\right)$.
Using $H(x,0)=0$, we note that $H\left(x,\nabla U/2\right)=-\psi^{*}\left(x,-\nabla U/2\right)$,
and thus 
\begin{equation}
L(x,\dot{x})=\psi(x,\dot{x})+\psi^{*}\left(x,-\frac{\nabla U}{2}\right)+\dot{x}.\frac{\nabla U}{2}.\label{eq:L_Psi}
\end{equation}
We recall that $L$ is non negative and that the most probable evolution
(the relaxation paths) are the paths $x(t)$ such that
\begin{equation}
\int_{0}^{T}\mbox{dt}\,L(x,\dot{x})=0,\label{eq:Relaxation_Caracterization}
\end{equation}
Comparing (\ref{eq:Relaxation_Caracterization}) and (\ref{eq:Gradient_Flow}),
and using (\ref{eq:L_Psi}) we immediately conclude that the relaxation
paths coincide with gradient flows with respect to the energy $E=U/2$
and dissipation function $\psi^{*}$.

For a path large deviation principle given by the Hamiltonian $H$
and the quasipotential $U$, we have thus proven that the relaxation
paths are gradient flows for the dissipation function $\psi^{*}$
given by (\pageref{eq:psi_star_H}) and the energy $E=U/2$.\\

Let us consider first the example of gradient diffusions 
\[
\mbox{d}X_{\epsilon}=-A\nabla U\mbox{d}t+\sqrt{2\epsilon}\sigma\mbox{d}W_{t}
\]
with $A\equiv\sigma\sigma^{T}$. One can compute the large deviation
principle with rate $\epsilon$ (see page \pageref{sec:Examples}).
The Hamiltonian is 
\[
H(x,p)=p.Ap-p.A\nabla U.
\]
It it easily checked that $U$ solves the stationary Hamilton-Jacobi
equation $H(x,\nabla U)=0$. With generic hypotheses, for instance
the hypotheses used in section \ref{subsec:A-sufficient-condition},
$U$ will be the quasipotential. It is easily checked that this dynamics
is time-reversible. From (\ref{eq:psi_star_H}), we compute the associated
dissipation function, and we find $\psi^{*}(x,p)=pAp$, with the relaxation
paths $\dot{X}=-A\nabla U$, as expected. 

\subsection{Dynamics with gradient-conservative structure and their relation
with path large deviations\label{subsec:Gradient-flow-for-1}}

Let us now define a time reversible-conservative structure for a path
large deviation. This definition is original. We consider a path large
deviation principle 
\begin{equation}
P\left[\left\{ X_{\epsilon}(t)\right\} _{0\leq t<T}=\left\{ X(t)\right\} _{0\leq t<T}\right]\underset{\epsilon\downarrow0}{\asymp}\exp\left(-\frac{\int_{0}^{T}\mbox{d}t\,L\left(X,\dot{X}\right)}{\epsilon}\right),\label{eq:Large deviations path-1-1}
\end{equation}
with $L\left(x,\dot{x}\right)=\sup_{p}\left\{ p\dot{x}-H(x,p)\right\} $,
and with quasipotential $U$ such that $H(x,\nabla U)=0$. We assume
that $H=H_{R}+H_{C}$ where $H_{R}$ the time reversible part of the
Hamiltonian which verifies the time reversal symmetry with respect
to the quasipotential $U$: for any $p$, $H_{R}\left(x,p+\nabla U\right)=H_{R}\left(x,-p\right)$,
and where $H_{C}$ is the conservative part of the Hamiltonian such
that for any $x$
\begin{equation}
\nabla U(x).\frac{\partial H_{C}}{\partial p}(x,0)=0.\label{eq:Conservation_Quasipotential}
\end{equation}
The relaxation path equation is 
\begin{equation}
\dot{x}=\frac{\partial H_{C}}{\partial p}(x,0)+\frac{\partial H_{R}}{\partial p}(x,0).\label{eq:Relaxation_grdient-transverse}
\end{equation}
We say that the vector field for the relaxation paths, $\frac{\partial H_{C}}{\partial p}(x,0)+\frac{\partial H_{R}}{\partial p}(x,0)$,
has a gradient/conservative decomposition, where $\frac{\partial H_{R}}{\partial p}(x,0)$
is the gradient part (the gradient of the quasipotential with respect
to a dissipative function) and $\frac{\partial H_{C}}{\partial p}(x,0)$
the conservative part (it conserves the quasipotential). Let me explain
a bit more.

Following the discussion of section \ref{subsec:Gradient-flow-for},
thanks to time reversal symmetry for $H_{R}$, we know that the dynamics
$\dot{x}=\frac{\partial H_{R}}{\partial p}(x,0)$ is a gradient flow
with dissipation function 
\[
\psi^{*}\left(x,p\right)=H_{R}\left(x,\frac{\nabla U}{2}+p\right)-H_{R}\left(x,\frac{\nabla U}{2}\right)
\]
and gradient flow energy $E=U/2$. Moreover equation $\dot{x}=\frac{\partial H_{C}}{\partial p}(x,0)$
conserves the value of the quasipotential thanks to the conservation
equation (\ref{eq:Conservation_Quasipotential}). As a consequence
the relaxation path equation systematically decreases the quasipotential,
through the gradient of the quasipotential with respect to the dissipative
function $\psi^{*}$ and the further mixing due the conservative part.
\\

Let us consider the example of diffusions with transverse decomposition:
\[
\mbox{d}X_{\epsilon}=-A\left(X_{\epsilon}\right)\nabla U\left(X_{\epsilon}\right)\mbox{d}t+G\left(X_{\epsilon}\right)\mbox{d}t+\sqrt{2\epsilon}\sigma\left(X_{\epsilon}\right)\mbox{d}W_{t}
\]
with $A\equiv\sigma\sigma^{T}$ and with the assumption that for any
$X$, $G(X).\nabla U(X)=0$ (transversality assumption). One can compute
the large deviation principle with rate $\epsilon$ (see page \pageref{sec:Examples}).
The Hamiltonian is 
\[
H(x,p)=p.Ap-p.A\nabla U+p.G.
\]
It it easily checked that $U$ solves the stationary Hamilton-Jacobi
equation $H(x,\nabla U)=0$, thanks to the transversality assumption.
With generic hypotheses, for instance the hypotheses used in section
\ref{subsec:A-sufficient-condition}, $U$ will be the quasipotential.
When $G\neq0$, in general this dynamics is \textbf{not} time-reversible.
As a consequence we do not expect this dynamics to be a gradient flow.
However it clearly has a time reversible-conservative structure, with
the definition given above. $H_{R}(x,p)=p.Ap-p.A\nabla U$ is the
reversible part of the Hamiltonian while $H_{C}(x,p)$ is the conservative
part. The transverse condition $G(X).\nabla U(X)=0$ is indeed a conservation
condition. As a consequence the relaxation path equation $\dot{X}=-A\nabla U+G$
has a gradient-conservative decomposition (in this case this is obvious).

We note that most equilibrium statistical mechanics stochastic processes
with Gaussian noises, described by stochastic differential equations,
for instance Hamiltonian dynamics in contact with a thermal bath,
have a gradient-conservative structure, where $G$ is the Hamiltonian
vector-field and $-A\nabla U$ the deterministic part of the interaction
with the thermal bath. Such a gradient-conservation structure is not
limited to equilibrium stochastic differential equations with Gaussian
noises. In the following section we discuss the gradient-conservation
structure of the Boltzmann equation.

\subsection{Gradient structure for the Boltzmann equation\label{subsec:Gradient-structure-for-Boltzmann}}

In section \ref{sec:The-reversible-Boltzmann} we have justified the
large deviation structure associated to the Boltzmann equation. The
large deviation principle (\ref{eq:Large_Deviation_Action-1}) is
an example of a large deviation principle (\ref{eq:Large deviations path-1})
discussed in section \ref{subsec:Gradient-flow-for-1}. The Hamiltonian
for the large deviations of the large deviations associated to the
Boltzmann equation is given by $H=H_{R}+H_{C}$ with $H_{R}$ given
by (\ref{eq:Hamiltonian-Collisions-1}) and $H_{C}$ given by (\ref{eq:Hamiltonian-Transport-1}).
We have justified in section \ref{sec:The-reversible-Boltzmann} that
$H_{R}$ has the time-reversal symmetry and that $H_{C}$ conserves
the entropy. As a consequence $H=H_{R}+H_{C}$ is a reversible-conservative
decomposition and the Boltzmann equation has a gradient-conservative
structure. The transport term of the Boltzmann equation is the conservative
one, while the collision term is the gradient of half the entropy
as a potential and a dissipation functional $\psi^{*}$ given by 
\[
\psi^{*}\left[f,p\right]=H_{R}\left(f,-\frac{1}{2}\frac{\delta S}{\delta f}+p\right)-H_{R}\left(f,-\frac{1}{2}\frac{\delta S}{\delta f}\right).
\]
This remark is very important from a physical point of view. This
gradient-conservative decomposition insures all the expected properties
of the entropy. Moreover the dissipation functional is the proper
local measure of the metric related to entropy changes. This remark
might also be of interest mathematically, as gradient-conservative
decomposition should prove extremely useful to define the proper functional
spaces and build the mathematical theory of the Boltzmann equation.\\

We conclude this section with a few simple remarks:
\begin{enumerate}
\item While we have an explicit formula for the Hamiltonian for the Boltzmann
equation, there is probably no explicit formula for the Lagrangian.
There is thus no explicit formula for neither $\psi$ the Legendre-Fenchel
conjugate of the dissipation function.
\item Some properties of $L$ can be found in the paper \cite{rezakhanlou1998large}.
\item It is easy to compute the dissipation function along a solution of
the Boltzmann equation that verifies $L\left[f,\dot{f}\right]=0$,
however this is just a subensemble of the space $\left[f,\dot{f}\right]$.
\item For the linearized Boltzmann equation close to the equilibrium, $H$
will be quadratic in $p$, and the expression for $L$ might be explicit
and the gradient structure might be explicit.
\end{enumerate}

\section{Conclusions and perspectives\label{sec:Conclusion-and-perspectives}}

In this paper, we have justified a large deviation principle for a
dilute gas in the Boltzmann-Grad limit. This large deviation principle
describes the probability of observing any evolution for the empirical
distribution in the position-velocity space. The solution of this
fundamental problem gives a specially clear perspective on the classical
irreversibility paradox. We have explained how the large deviation
action, Lagrangian and Hamiltonian, are natural consequences of the
Boltzmann molecular chaos hypothesis. Using this hypothesis, rather
than computing only the average evolution of the distribution function,
as Boltzmann did, we have computed its full distribution and obtained
the path large deviations. 

We guess that this large deviation functional will have profound implications.
First, the exercise described in this paper will probably be a classical
textbook one, because of its conceptual importance. Then it opens
the door to many questions. Can fluctuating hydrodynamics be obtained
from this large deviation action for the empirical distribution through
a hydrodynamic limit? Using this large deviation action, can we compute
the large deviations for the dynamics of macroscopic variables, for
instance for the evolution of a piston between two boxes? Can we compute
large deviation rate functions for current of particle, for dilute
gases flowing in between two boundaries in contact with thermal baths?
Could this large deviation principle be useful to study the property
and dynamics of non-smooth solutions to the Boltzmann and or Navier\textendash Stokes
equations? Are there physical applications of this large deviation
rate function? Could such an approach be generalized to other kinetic
equations?

Although the justification we gave, based on the chaotic hypothesis,
is extremely natural, a more precise justification would be welcomed.
Could one justify the same action from a hierarchy, for instance a
generalization of the BBGKY hierarchy? Would it be possible to give
a full mathematical proof of the validity of this large deviation
principle for short times? 

Finally we have also explained that this large deviation principle
implies a gradient-conservative structure for the Boltzmann equation.
We believe that this will have deep consequences in the future in
the mathematical study of the Boltzmann equation.

\section{Appendices}

\subsection{Large deviation rate functions from the infinitesimal generator of
a continuous time Markov process \label{sec:Large_Deviations_Generator}}

\subsubsection{Infinitesimal generator of a continuous time Markov process\label{subsec:Infinitesimal-generator-of}}

We recall the notion of the infinitesimal generator of a continuous
time Markov process. We consider the continuous time Markov processes
$\left\{ X(t)\right\} _{0\leq t\leq T}$, for instance $X(t)\in\mathbb{R}^{n}$.
The infinitesimal generator acts on the test function $\phi:\mathbb{R}^{n}\rightarrow\mathbb{R}$
and is defined by
\begin{equation}
G\left[\phi\right](x)=\lim_{t\downarrow0}\frac{\mathbb{E}_{x}\left[\phi(X(t))\right]-\mathbb{\phi}(x)}{t}.\label{eq:Infinitetisimal_Generator}
\end{equation}
For example, for a diffusion $\mbox{d}x=R(x)\mbox{d}t+\sqrt{2}\mbox{d}W_{t}$,
the infinitesimal generator is $G\left[\phi\right](x)=R(x)\nabla\phi+\Delta\phi$,
the adjoint of the Fokker\textendash Planck equation.

As an example, let us compute the infinitesimal generator for the
radioactive decay of a single particle, defined in section \ref{subsec:An-example:-large}.
If $X=1$ at time $t=0$, the probability that $X=1$ at time $t$,
for small $t$, is $1-\lambda t$ up to terms of order two in $t$.
The probability that $X=0$ at time $t$, for small $t$, is $\lambda t$,
up to terms of order two in $t$. If $X=0$ at time $t=0$, it remains
zero for all time. Then 
\[
G\left[\phi\right](1)=\lambda\left[\phi(0)-\phi(1)\right].
\]
and 
\[
G\left[\phi\right](0)=0.
\]
We can write
\[
G\left[\phi\right](x)=\lambda x(\phi(0)-\phi(1)).
\]
The generator is $(\phi(0)-\phi(1))$ the value of the function after
the jump minus its value before the jump multiplied by the jump rate
$\lambda$.

In the example of the radioactive decay, $X_{N}(t)$ (\ref{eq:X_N})
is also a continuous time Markov process. We can compute directly
its infinitesimal generator by studying all possible changes of the
variable $X_{N}$. We then obtain
\[
G_{N}\left[\phi\right](x)=N\lambda x\left[\phi\left(x-\frac{1}{N}\right)-\phi\left(x\right)\right],
\]
where $x=n/N$ with $n$ any integer number with $1\leq n\leq N$,
and $\phi$ is a real valued function on $\left[0,1\right]$. We also
have $G_{N}\left[\phi\right](0)=0$. The generator has on contribution
per jump: $(\phi(x-1/N)-\phi(x))$ the value of the function after
the jump minus its value before the jump multiplied by the jump rate
$N\lambda x$. The jump rate in this case is a single particle jump
rate $\lambda$ multiplied by the density $x$ multiplied by the total
particle number $N$.

\subsubsection{Heuristic derivation of the large deviation rate functions from the
infinitesimal generator of a continuous time Markov process \label{subsec:Heuristic-derivation}}

We give in this section a heuristic derivation of the relation between
(\ref{eq:H-Generator}), (\ref{eq:Lagrangian}), and (\ref{eq:Large deviations path}).

Let us consider trajectories $\left\{ X_{\epsilon}(t)\right\} _{0\leq t<\infty}$
starting at $x$. We denote $P_{t}\left(x,\dot{x}\right)$ the probability
that the Newton difference quotient $\frac{X(t)-x}{t}$ be equal to
$\dot{x}$ after a time $t$:
\begin{equation}
P_{t,\epsilon}\left(x,\dot{x}\right)\equiv\mathbb{E}_{x}\left[\delta\left(\frac{X_{\epsilon}(t)-x}{t}-\dot{x}\right)\right].\label{eq:P_Dt}
\end{equation}
Let us first assume that for small time $t$, $P_{t,\epsilon}$ verifies
the large deviation estimate 
\begin{equation}
P_{t,\epsilon}\left(x,\dot{x}\right)\underset{\epsilon\downarrow0}{\asymp}\exp\left(-\frac{tL\left(x,\dot{x}\right)}{\epsilon}\right)\label{eq:Lagrangien-Taux-D-Acroissement}
\end{equation}
(more precisely, we take first the limit $\epsilon\downarrow0$: $L\left(x,\dot{x}\right)=-\lim_{t\downarrow0}\lim_{\epsilon\downarrow0}\epsilon\log P_{t,\epsilon}\left(x,\dot{x}\right)/t$).
Then decomposing the path $X(t)$ in small subpaths, and using the
Markov property, we can construct a path integral and the large deviation
(\ref{eq:Large deviations path}) holds. It is thus sufficient to
prove the large deviation result (\ref{eq:Lagrangien-Taux-D-Acroissement})
holds in order to conclude that (\ref{eq:Large deviations path})
is true.

In order to assess the large deviation result (\ref{eq:Lagrangien-Taux-D-Acroissement}),
we can study the cumulant generating function of $P_{t,\epsilon}$.
A sufficient condition for (\ref{eq:Lagrangien-Taux-D-Acroissement})
to hold is then given by Gärtner\textendash Ellis theorem. If for
all $p$, the limit 
\begin{equation}
\begin{split}
H(x,p)&=\lim_{t\downarrow0}\lim_{\epsilon\downarrow0}\frac{\epsilon}{t}\log\mathbb{E}_{x}\left[\exp\left(\frac{t}{\epsilon}\frac{p.\left(X_{\epsilon}(t)-x\right)}{t}\right)\right]\\
&=\lim_{t\downarrow0}\lim_{\epsilon\downarrow0}\frac{\epsilon}{t}\log\left\{ \mathbb{E}_{x}\left[\exp\left(\frac{p.X_{\epsilon}(t)}{\epsilon}\right)\right]\exp\left(-\frac{p.x}{\epsilon}\right)\right\} \label{eq:H_Cumulant_Generating_Function}
\end{split}
\end{equation}
exists and $H$ is everywhere differentiable then (\ref{eq:Large deviations path})
will hold with $L$ given by (\ref{eq:Lagrangian}). Now, using the
definition of the infinitesimal generator (\ref{eq:Infinitetisimal_Generator})
we have

\begin{multline*}
\frac{1}{t}\log\left\{ \mathbb{E}_{x}\left[\exp\left(\frac{p.X_{\epsilon}(t)}{\epsilon}\right)\right]\exp\left(-\frac{px}{\epsilon}\right)\right\} \\
\begin{split}
&=\frac{1}{t}\log\left\{ 1+tG_{\epsilon}\left[\exp\left(\frac{px}{\epsilon}\right)\right]\exp\left(-\frac{px}{\epsilon}\right)+o(t)\right\} \\
&=G_{\epsilon}\left[\exp\left(\frac{px}{\epsilon}\right)\right]\exp\left(-\frac{px}{\epsilon}\right)+o(1).
\end{split}
\end{multline*}
Hence if the limit (\ref{eq:H-Generator}) exists then the large deviation
estimate (\ref{eq:Large deviations path}) holds.

\subsubsection{Examples\label{sec:Examples}}

The example of locally finitely indivisible processes is discussed
in Freidlin-Wentzell textbook. This includes the diffusion and Poisson
process cases discussed below. 

\paragraph{Diffusion with small noise.}

We consider the diffusion 
\[
\mbox{d}X_{\epsilon}=R(X_{\epsilon})\mbox{d}t+\sqrt{2\epsilon}\sigma(X_{\epsilon})\mbox{d}W_{t}
\]
where $X_{\epsilon}\in\mathbb{R}^{n}$, $R(.)$ is a vector field,
$\sigma$ a $n\times n$ matrix. We denote $a(x)\equiv\sigma(x)\sigma(x)^{T}$,
where $T$ stands for the transposition. The infinitesimal generator
is 
\[
G_{\epsilon}\left[\phi\right]=R.\nabla\phi+\epsilon a:\nabla\nabla\phi,
\]
where $:$ is the symbol for the contraction of two second order tensors.
Then it is easily checked that, from the definitions (\ref{eq:H-Generator})
and (\ref{eq:Lagrangian}),
\[
H(x,p)=p.ap+p.R.
\]
Then, whenever $a$ is invertible,
\[
L(x,\dot{x})=\frac{1}{4}\left(\dot{x}-R\right).a^{-1}\left(\dot{x}-R\right).
\]
$H$ and $L$ are the classical Hamiltonian and Lagrangian for a diffusion
with small noise.

\paragraph{Poisson process}

The case of a Poisson process is discussed in the book of Freidlin\textendash Wentzell.
This textbook considers a single Poisson process, rescaled in order
to have an infinitesimal generator that fits with the asymptotics
leading to a large deviation estimate, as in equation (\ref{eq:H-Generator}).
We rather consider $N$ independent Poisson processes $\left\{ x_{n}(t)\right\} _{1\leq n\leq N}$
for which we will look at the large deviations for their empirical
average. This case is more in line with what will be needed in this
paper. The value $x_{n}$ of each of these Poisson processes is increased
by $1$ at a rate $1$ (the probability of $x_{n}$ to increase by
a jump equal to one during an infinitesimal time interval $dt$ is
$dt$).

We consider the average
\[
X_{N}(t)=\frac{1}{N}\sum_{n=1}^{N}x_{n}(t).
\]
During an infinitesimal interval $dt$, the probability for $X_{N}(t)$
to increases by an amount $1/N$ is $Ndt$. The infinitesimal generator
is thus 
\[
G_{N}\left[\phi\right](x)=N\left[\phi\left(x+\frac{1}{N}\right)-\phi\left(x\right)\right].
\]
Using (\ref{eq:H-Generator}) with $\epsilon=1/N$, we deduce that
the process $\left\{ X_{N}(t)\right\} $ verifies a large deviation
principle with an action characterized by the Hamiltonian
\[
H(x,p)=\exp(p)-1.
\]

\subsection{Quasipotential, relaxation paths, fluctuation paths, and conservation
laws\label{subsec:Quasipotential,-relaxation-paths}}

\subsubsection{Some properties of the Lagrangian and of the Hamiltonian\label{subsec:Some-properties-of}}

$L$ is a large deviation rate function for the variable $\dot{x}$.
By definition, a large deviation rate function has zero as its minimum
value. We thus have 
\begin{equation}
\inf_{\dot{x}}L(x,\dot{x})=0=L\left(x,R(x)\right),\label{eq:Definition_R}
\end{equation}
where for the second equality we assume that the infimum is achieved
at $\dot{x}=R(x)$. We also have 
\begin{equation}
L(x,\dot{x})\geq0.\label{eq:L_Positif}
\end{equation}

From the definition of the Hamiltonian H as a rescaled cumulant generating
function (\ref{eq:H_Cumulant_Generating_Function}), we can conclude
that for any $x,$ $H$ is a convex function of the variable $p$
and that 
\begin{equation}
H(x,0)=0.\label{eq:H_0}
\end{equation}
The Legendre\textendash Fenchel relation between $L$ and $H$ (\ref{eq:Lagrangian})
implies that for any $x$ and $p$ 
\begin{equation}
p\dot{x}\leq L(x,\dot{x})+H(x,p)\label{eq:Convexity_Inequality-1}
\end{equation}
from which, using (\ref{eq:H_0}) we verify again (\ref{eq:L_Positif}).

\subsubsection{Relaxation paths\label{subsec:Relaxation-paths}}

The relaxation paths $X_{r}(t,x)$ are the most probable paths of
the dynamics described by the action (\ref{eq:Action}), starting
from a state $x$ at time $t=0$. They thus minimize the action. From
the definition of $R$ (\ref{eq:Definition_R}), as $L\geq0$ and
$L\left(x,R(x)\right)=0$, relaxation paths thus solve
\begin{equation}
\dot{X}_{r}=R(X_{r}),\label{eq:Relaxation_Paths}
\end{equation}
with the initial condition $X_{r}(0,x)=x$.

Moreover looking at the condition for the stationarity of the variational
problem $0=L(X,R(X))=\sup_{p}\left[p.R(X)-H\left(X,p\right)\right]$,
we conclude that the optimal is achieved for $p=0$ and that
\begin{equation}
R(X_{r})=\frac{\partial H}{\partial p}(X_{r},0).\label{eq:R}
\end{equation}

In the following, in order to keep the discussion simple, we assume
that the relaxation dynamics has a single global point attractor $x_{0}$,
with $R(x_{0})=0$. The generalization to multiple attractors or to
other types of attractors could be considered following the classical
discussion (see for instance \cite{Freidlin_Wentzel_1984_book}).
As we will see, this hypothesis will be verified for the Boltzmann
equation.

\subsubsection{Quasipotential\label{subsec:Quasipotential}}

We consider now the stationary distribution $P_{s}$ of the processes
$X_{\varepsilon}$ which dynamics follows the large deviation principle
(\ref{eq:Large deviations path}). We assume that the stationary distribution
also follows a large deviation principle: 
\begin{equation}
P_{s}(x)\equiv\mathbb{E}\left[\delta\left(X_{\epsilon}-x\right)\right]\underset{\epsilon\downarrow0}{\asymp}\exp\left(-\frac{U(x)}{\epsilon}\right),\label{eq:Quasipotential_Definition}
\end{equation}
where $U$ is called the quasipotential. In the case when the relaxation
dynamics has a single global attractor $x_{0}$, the quasipotential
is characterized by the variational problem
\begin{equation}
\begin{split}
U(x)&=\inf_{\left\{ X(t)\left|X(-\infty)=x_{0}\,\,\mbox{and}\,\,X(0)=x\right.\right\} }\int_{-\infty}^{0}\mbox{d}t\,L(X,\dot{X})\\
&=\inf_{\left\{ X(t),P(t)\left|X(-\infty)=x_{0}\,\,\mbox{and}\,\,X(0)=x\right.\right\} }\int_{-\infty}^{0}\mbox{d}t\,\left[P\dot{X}-H(X,P)\right].\label{eq:Quasipotential_Variational_Problem}
\end{split}
\end{equation}
It is a classical result, that can be found for instance in any textbook
of classical mechanics, that the minimum of a variational problem
with a Lagrangian solves a Hamilton\textendash Jacobi equation. Then
the quasipotential $U$ solves the stationary Hamilton\textendash Jacobi
equation 
\begin{equation}
H(x,\nabla U)=0.\label{eq:Hamilton_Jacobi}
\end{equation}

\subsubsection{Fluctuation paths\label{subsec:Fluctuation-paths}}

The fluctuation paths are the minimizers of the quasipotential variational
problem (\ref{eq:Quasipotential_Variational_Problem}). They are very
important as they describe the most probable path starting from the
attractor $x_{0}$ and leading to a fluctuation $x$.

The fluctuation paths define a flow parametrized by $x,$ that we
denote $X_{f}(t,x)$ (the path evolution) and $P_{f}(t,x)$ (the conjugated
momentum evolution). They verify the Euler-Lagrange equations 
\begin{equation}
\left\{ \begin{array}{ccc}
\dot{X_{f}} & = & \frac{\partial H}{\partial p}\left(X_{f},P_{f}\right)\\
\dot{P}_{f} & = & -\frac{\partial H}{\partial x}\left(X_{f},P_{f}\right),
\end{array}\right.\label{eq:Hamilton_Equation}
\end{equation}
with the boundary conditions $X_{f}(-\infty,x)=x_{0}$ and $X_{f}(0,x)=x$.
As any fluctuation path converges to $x_{0}$ as $t\downarrow-\infty$,
we have $R(x_{0})=\frac{\partial H}{\partial p}(x_{0},0)=0$. As $H$
is a convex function of the variable $p$, the equation $\frac{\partial H}{\partial P}(x_{0},p)=0$
can have at most one root, from which we deduce that $\lim{}_{t\downarrow-\infty}P_{f}(t,x)=0$.
Moreover, Hamilton's equations (\ref{eq:Hamilton_Equation}) conserve
the Hamiltonian $H$ along their dynamics. From the value of $H$
for $t\downarrow-\infty$, we deduce that along the fluctuation paths
$H(X_{f},P_{f})=0$. From the variational characterization of the
quasipotential (\ref{eq:Quasipotential_Variational_Problem}), we
then deduce that $U(x)=\int_{-\infty}^{0}\mbox{d}t\,P_{f}(t,x)\dot{X}_{f}(t,x)$.
It then follows that $\nabla U(x)=P_{f}(0,x)$. Using the flow property,
it is clear that this relation is valid all along the fluctuation
paths. Then for any $x$ and $t$ 
\[
\nabla U(X_{f}(t,x))=P_{f}(t,x).
\]
Using this result and (\ref{eq:Hamilton_Equation}), we deduce that
the fluctuation paths solve the first order equation
\begin{equation}
\dot{X}_{f}=F(X_{f})\equiv\frac{\partial H}{\partial p}\left(X_{f},\nabla U(X_{f})\right),\label{eq:Equation_Fluctuation_Paths}
\end{equation}
where the second equality defines the fluctuation path vector field
$F$.

\subsubsection{Decay (resp. increase) of the quasi potential along the relaxation(
resp. fluctuation) paths\label{subsec:Decay/increase-of-the}}

We now prove that the value of $U$ characterizes the relaxation towards
the attractor $x_{0}$: any relaxation path decreases $U$ monotonously.
Indeed, from (\ref{eq:Relaxation_Paths}) and (\ref{eq:R}), we have
\begin{equation*}
\begin{split}
\frac{\mbox{d}U}{\mbox{d}t}(X_{r})&=\frac{\partial H}{\partial p}\left(X_{r},0\right).\nabla U(X_{r})\\
&=H(X_{r},0)-H(X_{r},\nabla U(X_{r}))+\frac{\partial H}{\partial p}\left(X_{r},0\right).\nabla U(X_{r})\leq0
\end{split}
\end{equation*}
where we have used (\ref{eq:H_0}) and the Hamilton\textendash Jacobi
equation (\ref{eq:Hamilton_Jacobi}) to write the second equality.
The inequality is a consequence of the convexity of $H$ with respect
to its second variable. In case of strict convexity, which will be
often the case, the equality holds if and only if $\nabla U(X_{r})=0$. 

Moreover, the condition: for any $\alpha\in[0,1]$
\begin{equation}
\left(\nabla U\right)^{T}\frac{\partial^{2}H}{\partial p^{2}}\left(X_{r},\alpha\nabla U\right)\nabla U\geq CU\label{eq:Condition_Dissipation}
\end{equation}
implies a convergence to equilibrium faster than $\mbox{e}^{-Ct}$.
The condition that the quasipotential is uniformly convex in the norm
of the second variation of $H$: for any $p$
\begin{equation}
p^{T}\frac{\partial^{2}H}{\partial p^{2}}\left(x,\alpha\nabla U\right)\mbox{Hess}U(x)p\geq Cp^{T}p\label{eq:Positivite_Hessienne_NormeH}
\end{equation}
implies (\ref{eq:Condition_Dissipation}). In the case of the sum
of $N$ independent particles, where each follows a diffusion, the
second variations of $H$ are the Wasserstein distance and the condition
(\ref{eq:Positivite_Hessienne_NormeH}) is a log-Sobolev inequality.\\

We now prove similarly that $U$ increases monotonously along the
fluctuation paths. Using (\ref{eq:Equation_Fluctuation_Paths}), we
have
\begin{equation*}
\begin{split}
\frac{\mbox{d}U}{\mbox{d}t}(X_{f})&=\frac{\partial H}{\partial P}\left(X_{f},\nabla U(X_{f})\right).\nabla U(X_{f})\\
&=H(X_{f},0)-H(X_{f},\nabla U(X_{f}))+\frac{\partial H}{\partial P}\left(X_{f},\nabla U(X_{f})\right).\nabla U(X_{f})\geq0,
\end{split}
\end{equation*}
where the second equality is a consequence of the Hamilton\textendash Jacobi
equation (\ref{eq:Hamilton_Jacobi}) and of (\ref{eq:H_0}), and the
inequality is again a consequence of the convexity of $H$ with respect
to its second argument. Again if $H$ is strictly convex the inequality
is strict whenever $\nabla U(X_{r})\neq0$.

\subsubsection{Conservation laws\label{subsec:Conservation-law-Hamiltonien}}

It may happen that the stochastic process has a conservation law $C$:
for any $\epsilon$ and $t$, $C\left(X_{\epsilon}(t)\right)=C_{0}$.
Then, $\dot{X}_{\epsilon}(t).\frac{\partial C}{\partial x}(X_{\epsilon}(t))=0$.
As a consequence, from the definition of the Lagrangian (\ref{eq:Lagrangien-Taux-D-Acroissement}),
we deduce that 
\[
L(x,\dot{x})=+\infty\,\,\,\mbox{if}\,\,\,\dot{x}.\frac{\partial C}{\partial x}(x)\neq0.
\]
At the level of the Hamiltonian, using the Legendre\textendash Fenchel
transform (\ref{eq:Lagrangian}), we conclude that the conservation
law translates to the continuous symmetry property
\begin{equation}
\mbox{for any}\,\,x,p\,\,\mbox{and}\,\,\alpha,\,\,\,H\left(x,p+\alpha\frac{\partial C}{\partial x}\right)=H(x,p)\label{eq:Conservation_Law_H_0}
\end{equation}
or equivalently
\begin{equation}
\mbox{for any}\,\,x\,\,\mbox{and}\,\,p,\,\,\,\frac{\partial H}{\partial p}\left(x,p\right).\frac{\partial C}{\partial x}(x)=0.\label{eq:Conservation_Law_H}
\end{equation}
Then as a function of its second variable, $H(x,.)$ is flat in the
direction $\frac{\partial C}{\partial x}$. 

As far as the Hamilton\textendash Jacobi equation is concerned, this
means that only the projection of the gradient of $U$ on the orthogonal
of $\nabla C$ matters.

\subsection{Time reversal symmetry and detailed balance\label{subsec:Detailed-balance-general}}

\subsubsection{Detailed balance\label{subsec:Detailed-balance}}

If the Markov process is time reversible, or equivalently if it verifies a detailed balance condition, this implies a time reversal symmetry for the path large deviation estimate (\ref{eq:Large deviations path}),
or equivalently the action (\ref{eq:Action}). We explain this point in this section. 

A stationary continuous time Markov process is said to be time reversible if its backward and forward histories have the same probabilities. We consider the transition probability
$P_{T}$ for the Markov process ($P_{T}(y;x)$ is the transition probability from the state
$x$ towards the state $y$). If for any states $x$ and $y$, 
\begin{equation}
P_{T}(y;x)P_{S}(x)=P_{T}(x;y)P_{S}(y),\label{eq:Detailed_Balance}
\end{equation}
we say that the process verifies a detailed balance property with
respect to the distribution $P_{s}$. It is then very easily checked
that $P_{S}$ is a stationary distribution of the Markov process. The detailed balance condition is a necessary and sufficient condition for the Markov process to be time reversible. Another characterization of the time-reversibility of the process is that the infinitesimal generator of the time reversed process is identical to the infinitesimal generator of the initial process.

If for any $\epsilon$ the process $\left\{ X_{\epsilon}\right\} $
verifies a detailed balance property, then the large deviation dynamics
will inherit this symmetry property. However, the converse is not
necessarily true, the detailed balance property can hold at the level
of the large deviations dynamics without holding at the level of the
process $\left\{ X_{\epsilon}\right\} $.

For the process we are interested in, the condition for detailed balance
can be written 
\[
P_{\Delta t,\epsilon}\left(x,\dot{x}\right)P_{S}(x)\underset{\Delta t\rightarrow0}{\sim}P_{\Delta t,\epsilon}\left(x+\Delta t\dot{x},-\dot{x}\right)P_{S}(x+\Delta t\dot{x}),
\]
where $P_{\Delta t,\epsilon}$ is defined by (\ref{eq:P_Dt}). Using
the large deviation estimates (\ref{eq:Lagrangien-Taux-D-Acroissement})
and (\ref{eq:Quasipotential_Definition}) evaluated for small $\Delta t$,
the detailed balance condition writes: for any $x$ and $\dot{x}$
\begin{equation}
L(x,\dot{x})-L(x,-\dot{x})=\dot{x}.\nabla U.\label{eq:Detailed_Balance_Lagrangian}
\end{equation}
Using the Legendre\textendash Fenchel relations between $H$ and $L$
(\ref{eq:Lagrangian}), this detailed balance condition writes: for
any $x$ and $p$
\begin{equation}
H\left(x,-p\right)=H\left(x,p+\nabla U\right).\label{eq:Detailed_Balance_Hamiltonian}
\end{equation}
If $H$ and $U$ verify the detailed balance condition (\ref{eq:Detailed_Balance_Hamiltonian}),
using $H(x,0)=0$ (eq. (\ref{eq:H_0})), we easily deduce that $H\left(x,\nabla U\right)=0$
which is the Hamilton-Jacobi equation. With some further conditions
on $U$, see for instance section \ref{subsec:A-sufficient-condition},
we may conclude that $U$ is the quasipotential.

Moreover, if detailed balance is verified, then one expects to observe
the time reversal symmetry at the level of the relaxation and fluctuation
paths. Indeed from (\ref{eq:Detailed_Balance_Hamiltonian}), we easily
derive $R(x)=\frac{\partial H}{\partial x}\left(x,0\right)=-\frac{\partial H}{\partial x}\left(x,\nabla U\right)=-F(x)$.
We thus conclude that for Hamiltonians with detailed balance relation,
the fluctuation paths are the time reversed of the relaxation paths.

\subsubsection{Generalized detailed balance\label{subsec:Generalized-detailed-balance}}

For most physical systems the notion of time reversibility has to be extended, for instance in order to take into account that the velocity sign has to be changed in systems with inertia, or other fields have to be modified in the time-reversal symmetry. This is true for the time-reversal symmetry of dynamical systems, for instance of mechanical systems described by Hamiltonian equations, but also for the time-reversal symmetry of Markov processes. Such a generalized definition of time reversal symmetry is classical both in the physics and the mathematics literature, see for instance \cite{Gardiner_1994_Book_Stochastic}. 

We consider a map $I$ from the state space to itself. We assume
that $I$ is an involution ($I^{2}=Id$) and that $I$ is self adjoint
for the canonical scalar product: for any $x$ and $y$, $I(x).y=x.I(y)$. A continuous time Markov process is said to be time-reversal symmetric in the generalized sense if its backward histories with the application of $I$ and its forward histories have the same probabilities.
If the distribution $P_{S}$ is $I$-symmetric (for any $x$ $P_{S}\left(I(x)\right)=P_{S}(x)$)
and if for any states $x$ and $y$, 
\begin{equation}
P_{T}(y;x)P_{S}(x)=P_{T}\left(I(x);I(y)\right)P_{S}\left(I(y)\right),\label{eq:Generalized_Detailed_Balance}
\end{equation}
we say that the process verifies a generalized detailed balance property
with respect to the distribution $P_{s}$ and the symmetry $I$. It
is then very easily checked that $P_{S}$ is a stationary distribution
of the Markov process. The generalized detailed balance condition is a necessary and sufficient condition for the Markov process to be time reversible in the generalized sense. Another characterization of the time-reversibility of the process in the generalized sense is that the infinitesimal generator of the time reversed process is identical to the generator of the initial process up to application of the involution $I$. 

Then, the discussion of section \ref{subsec:Detailed-balance} easily
generalizes. The conditions of generalized detailed balance at the
level of large deviation read $U\left(I(x)\right)=U(x)$ and
\[
L(x,\dot{x})-L(x,-I\left[\dot{x}\right])=I\left[\dot{x}\right].\nabla U
\]
or equivalently 
\[
H\left(I\left[x\right],-I\left[p\right]\right)=H\left(x,p+\nabla U\right).
\]
If a generalized detailed balance is verified, then the quasipotential
solves the Hamilton\textendash Jacobi equation and the fluctuation
paths are the time reversed of the fluctuation paths composed with
the symmetry $I$: $F(x)=-R\left(I\left(x\right)\right)$.

\subsection{A sufficient condition for $U$ to be the quasipotential\label{subsec:A-sufficient-condition}}

We know that if $U$ is the quasipotential then it solves the Hamilton\textendash Jacobi
equation $H\left(x,\nabla U\right)=0$. The converse is not necessarily
true. For instance $U=0$ solves the Hamilton\textendash Jacobi equation
but is not the quasipotential. 

We give a sufficient condition for $U$ to be the quasipotential,
in the simple case when $U$ has a unique global minimum $x_{0}$.\\

If $V$ solves the Hamilton\textendash Jacobi equation and $V$ has
a single minimum $x_{0}$ with $V(x_{0})=0$, if moreover for any
$x$ the solution of the reverse fluctuation path dynamics $\dot{X}=-F(X)=-\frac{\partial H}{\partial p}\left(X,\nabla V(X)\right)$,
with $X(t=0)=x$, converges to $x_{0}$ for large times, then $V$
is the quasipotential. We give now a simple proof. 

From the definition of $L$ (\ref{eq:Lagrangian}), we have for any
$X$ and $\dot{X}$, $L\left(X,\dot{X}\right)\geq\dot{X}\nabla V(X)-H\left(X,\nabla V(X)\right)$.
Hence using that $V$ solves the Hamilton, Jacobi equation ($H\left(x,\nabla V\right)=0$),
we obtain that for any $X$ such that $X(0)=x$ and $X(-\infty)=x_{0}$
\[
\int_{-\infty}^{0}\mbox{d}t\,L\left(X,\dot{X}\right)\geq\int_{-\infty}^{0}\mbox{d}t\,\dot{X}\nabla V(X)=V(x).
\]
Hence, using the characterization of the quasipotential (\ref{eq:Quasipotential_Variational_Problem}),
we get $U(x)\geq V(x)$. 

Moreover, from the definition of $L$ (\ref{eq:Lagrangian}), for
any $x$ and $p$ we have
\[
L\left(x,\frac{\partial H}{\partial p}(x,p)\right)=p\frac{\partial H}{\partial p}(x,p)-H(x,p).
\]
If we apply this formula to the fluctuation path that verifies $\dot{X_{f}}=\frac{\partial H}{\partial p}\left(X_{f},\nabla V(X_{f})\right)$,
with $p=\nabla U$, using moreover $H(x,\nabla V(x))=0$, we get 
\[
\int_{-\infty}^{0}\mbox{d}t\,L\left(X_{f},\dot{X}_{f}\right)=\int_{-\infty}^{0}\mbox{d}t\,\dot{X}_{f}\nabla V(X_{f})=V(x).
\]
Hence $U(x)\leq V(x)$. We thus conclude that $V$ is the quasipotential.

\subsection{The infinitesimal generator for the free transport.\label{subsec:The-infinitesimal-generator}}

We consider $N$ particles that undergo free transport. Each particle
$1\leq n\leq N$ has a position $\mathbf{r}_{n}(t)$ and a velocity
$\mathbf{v}_{n}(t)$. Then the empirical distribution $f$ verifies
the equation 
\[
\frac{\partial f}{\partial t}=-\mathbf{v}.\frac{\partial f}{\partial\mathbf{r}}
\]
Let us consider a $\phi$ functional of $f$. Then $\phi$ evolves
according to 
\[
\frac{\mbox{d}\phi}{\mbox{d}t}=\int\mbox{d}\mathbf{r}\mbox{d}\mathbf{v}\,\frac{\partial f}{\partial t}(\mathbf{r},\mathbf{v})\frac{\delta\phi}{\delta f(\mathbf{r},\mathbf{v})}=-\int\mbox{d}\mathbf{r}\mbox{d}\mathbf{v}\,\mathbf{v}.\frac{\partial f}{\partial\mathbf{r}}(\mathbf{r},\mathbf{v})\frac{\delta\phi}{\delta f(\mathbf{r},\mathbf{v})}.
\]
Then the infinitesimal generator of the free transport is
\[
G\left[\phi\right]=-\int\mbox{d}\mathbf{r}\mbox{d}\mathbf{v}\,\mathbf{v}.\frac{\partial f}{\partial\mathbf{r}}(\mathbf{r},\mathbf{v})\frac{\delta\phi}{\delta f(\mathbf{r},\mathbf{v})}.
\]
If $\phi=\mbox{e}^{\frac{\int\text{d}\mathbf{r}\text{d}\mathbf{v}\,pf}{\epsilon}}$,
then 
\[
\frac{\delta\phi}{\delta f(\mathbf{r},\mathbf{v})}=\frac{p(\mathbf{r},\mathbf{v})}{\epsilon}\mbox{e}^{\frac{\int d\mathbf{r}_{1}d\mathbf{v}_{1}\,pf}{\epsilon}}
\]
and 
\[
\epsilon G\left[\mbox{e}^{\frac{\int d\mathbf{r}d\mathbf{v}\,pf}{\epsilon}}\right]\mbox{e}^{-\frac{\int\text{d}\mathbf{r}\text{d}\mathbf{v}\,pf}{\epsilon}}=-\int\mbox{d}\mathbf{r}\mbox{d}\mathbf{v}\,p(\mathbf{r},\mathbf{v})\mathbf{v}.\frac{\partial f}{\partial\mathbf{r}}(\mathbf{r},\mathbf{v}).
\]

\begin{acknowledgements}
This work has been initiated following discussions with L. Saint Raymond.
I thank her for very fruitful discussions. I thank C. Villani for
pointing me to the works of F. Rezakhanlou, in 2015, after I derived
this large deviation principle from a chaotic hypothesis. I thank
G. Eyink, O. Feliachi, J. Reygner and E. Woillez for comments on this manuscript. The research
leading to these results has received funding from the European Research
Council under the European Union\textquoteright s seventh Framework
Programme (FP7/2007-2013 Grant Agreement No. 616811. In its last stage,
this work was supported by a Subagreement from the Johns Hopkins University with funds provided by Grant No. 663054 from Simons Foundation. Its contents are solely the responsibility of the authors and do not necessarily represent the official views of Simons Foundation or the Johns Hopkins University.
\end{acknowledgements}

%
%

\bibliographystyle{spmpsci}      
\bibliography{boltzmann}


\end{document}